\documentclass[12pt]{article}

\usepackage{jheppub}
\usepackage{amsbsy,amsfonts,amsmath,amssymb}
\usepackage{cleveref} 
\usepackage[g]{esvect} 

\crefname{equation}{}{}
\crefname{chapter}{Chapter}{Chapters}
\crefname{section}{Section}{Sections}
\crefname{subsection}{Subsection}{Subsections}
\crefname{subsubsection}{Subsubsection}{Subsubsections}
\crefname{figure}{Figure}{Figures}
\crefname{table}{Table}{Tables}
\crefname{appendix}{Appendix}{Appendices}

\allowdisplaybreaks

\begin{document}

\title{Compactified AdS black holes, Chamblin-Reall background, and their dual non-conformal relativistic fluids}

\author[]{Chao Wu and Yanqi Wang}

\affiliation[]{School of Physics and Optoelectronic Engineering, Anhui University, Hefei 230601, China}

\emailAdd{chaowu86@outlook.com, wangyanqi0@gmail.com}

\abstract{The Chamblin-Reall background is a static solution of Einstein gravity coupled with a background scalar field and a dynamical domain wall, with the potential of the scalar field being of Liouville type. It can be got by dimensionally reducing a higher dimensional background with a constant potential. Compactified AdS black holes are black hole backgrounds constructed by wrapping one or more spatial directions of a higher dimensional AdS black hole on a torus and then integrating them out. The compactified AdS black hole background is asymptotically flat, non-conformal, and of Chamblin-Reall type. In this work, we derive all the 7 dynamical second-order transport coefficients for the relativistic fluids dual to compactified AdS black holes of various dimensions via fluid/gravity correspondence. Through this work, we achieve three main goals: (1) We prove that all the gravitational backgrounds that can be used to extract analytical results for second-order transport coefficients hitherto are all Chamblin-Reall type backgrounds. (2) We generalize the results in previous studies on the second-order transport coefficients of the relativistic fluids dual to 5-dimensional Chamblin-Reall model into general dimensions. (3) We offer a thorough study on the Kanitscheider-Skenderis proposal and find its physical accounts.}


\arxivnumber{2111.04091}

\maketitle

\section{Introduction and motivation}

The idea of this work originates from our calculation of the second-order transport coefficients of compactified Dp-brane \cite{Wu2012}. We know that the near-extremal black D3-brane will be trivially dimensional reduced to AdS$_5$ black hole after integrating out the 5-dimensional sphere, whose dual relativistic fluid is conformal and has been thoroughly studied by two holographic methods: the Minkowskian AdS/CFT correspondence \cite{Son0205,Policastro0205,Policastro0210,Baier0712,Arnold1105} and the fluid/gravity correspondence \cite{Bhattacharyya0712,Bhattacharyya0803}. In the case of compactified Dp-brane, both the $(8-p)$-dimensional sphere and a number of $q$ directions of Dp-brane will be reduced. In the case $p=3$, the reduction in the directions of the brane equals to reducing the spatial directions of the AdS$_5$ black hole, leaving us a black hole background of lower dimensions. We call the black hole backgrounds that are got by dimensionally reduced from a higher dimensional AdS black hole the compactified AdS black holes. In this work, we will show that the compactified AdS black holes, together with the backgrounds of compactified Dp-brane, are both Chamblin-Reall type backgrounds. And the compactified AdS black hole is non-conformal, asymptotically flat in any dimensions. The dynamical second-order transport coefficients of the strongly coupled, non-conformal relativistic fluids dual to compactified AdS black holes will also be offered by direct calculation.

In the literature, non-conformality is usually achieved in two ways. The first is to deform the AdS$_5$ black hole by manually adding a background scalar field. References along this line are \cite{Gubser0804PRL,Gubser0804PRD,Gubser0806,Finazzo1412,Attems1603,Kleinert1610,Li1411}. According to the analytic form or the way to solve the scalar potential, these works may be divided into three classes:
\begin{itemize}
\item
The Gubser model \cite{Gubser0804PRL,Gubser0804PRD,Gubser0806,Finazzo1412}, in which the scalar potential is a combination of a hyperbolic cosine and a 6 order polynomial with only even powers of the scalar field.
\item
The holographic renormalization group (RG) flow model \cite{Attems1603,Kleinert1610}, whose scalar potentials are even-powered 8 or 6 order polynomials.
\item
The dynamical holographic QCD model \cite{Li1411}. It is different from the former two in that the scalar field itself is set manually, and the potential needs to be determined through the equations of motion (EOMs).
\end{itemize}
The most salient feature for the models of scalar-deformed AdS$_5$ black holes is that either the potential or the scalar field needs to be set by hand. So those models all contain free parameters adjusted by hand with the solutions and final results are all numerical.

The second way to enter the non-conformal regime is to use a non-conformal background from the start. The non-conformal backgrounds can also be divided into three classes:
\begin{itemize}
\item
The supergravity background \cite{Buchel0311,Buchel0406200,Benincasa0507,Buchel0812,Buchel0908,Buchel1110,Buchel1503,
Bigazzi0909,Bigazzi0912,Buchel0509,Buchel0903}. There are three backgrounds in this class. The salient similarity shared by all of them is that the background metrics are all asymptotically AdS$_5$ and they can not be solved both exactly and analytically. The existing analytical solutions for these backgrounds are all perturbative---causing the results are not formulated in full non-conformal forms. The details will be specified in a later section.
These models can be classified by the independent background scalar fields they have. The first is the gravity dual of $\mathcal N=2^*$ supersymmetric gauge theory \cite{Buchel0311,Buchel0406200,Benincasa0507,Buchel0812,Buchel0908,Buchel1110,Buchel1503}, which has two background scalar fields. It is interesting to note that the author of \cite{Buchel1110} wraps the 5-dimensional bulk gravity of $\mathcal N=2^*$ theory on a 2-dimensional compact manifold and finds that the bulk viscosity extracted from the compactified bulk gravity dual to $\mathcal N=2^*$ theory violates the lower bound of bulk viscosity proposed in \cite{Buchel0708}.
The second is the background of near-extremal D3-branes with D7-branes' back reaction \cite{Bigazzi0909,Bigazzi0912}. The 5-dimensional reduced theory of this model is of Chamblin-Reall type with 3 independent scalars.
The last is the Klebanov-Tseytlin background, which is the gravity dual of the cascading gauge theory \cite{Buchel0509,Buchel0903}. Its 5-dimensional reduced theory has 4 independent background scalars.
\item
The 5-dimensional Chamblin-Reall background \cite{Gubser0804PRD,Gubser0806,Bigazzi1006,Kleinert1610}. Chamblin-Reall background \cite{Chamblin9903} is a classical gravitational solution that Einstein gravity coupled with a background scalar field and a dynamical domain wall. It can exist in any dimension with the potential of the scalar field the Liouville potential, i.e. a function in the form of the exponential of that scalar field. According to  \cite{Chamblin9903}, Chamblin-Reall background can be got by dimensional reduction from a higher dimensional Einstein gravity with a constant potential. All the studies we mentioned here, i.e. \cite{Gubser0804PRD,Gubser0806,Bigazzi1006,Kleinert1610} use the 5-dimensional Chamblin-Reall background which originates from a $(p+2)$-dimensional Einstein gravity with negative cosmological constant with $p-3$ of its spatial directions compactified on a torus of the same dimensions\footnote{Here we change the convention of the dimensions here. What is used in the original studies is $D=p+1$.}, i.e. $\text{AdS}_{p+2} \to \text{CR}_5 \times \mathbf T^{p-3}$. What's more, \cite{Gubser0804PRD,Gubser0806} study the first-order transport properties and \cite{Bigazzi1006,Kleinert1610} calculate the second-order transport coefficients for the boundary fluid of $\text{CR}_5$.
\item
The brane background
\cite{Parnachev0506,Benincasa0605,Mas0703,Natsuume0807,Springer0810,Springer0902,Kanitscheider0901,
David0901,Wu1508,Wu1604,Wu1608,Wu1807,Wu2012}. Therein, \cite{Parnachev0506} is about the NS5-brane and \cite{Benincasa0605,Mas0703,Natsuume0807,Springer0810,Springer0902,Kanitscheider0901,David0901,
Wu1508,Wu1604,Wu1608,Wu1807,Wu2012} are about the Dp- or compactified Dp-brane backgrounds. We will show that the reduced metric after integrating out the compact directions on the compactified Dp-brane is also of Chamblin-Reall type.
\end{itemize}
The most obvious feature about the supergravity background is that the results are formulated in a form of leading order non-conformal corrections to the conformal limit via some conformal breaking parameters. Whereas the results of Chamblin-Reall and brane backgrounds are completely analytic, exact, and in full non-conformal forms. Another important observation is that there is no free parameter in the results of the Chamblin-Reall and brane backgrounds.

In this paper, we generalize the studies on 5-dimensional Chamblin-Reall background by $\text{AdS}_{p+2} \to \text{CR}_{p-q+2} \times \mathbf T^q$ with both $p,\ q$ of general number as long as $p\geq 2$ and $1 \leq q \leq p-1$. The reason is that we need to leave at least one spatial direction\footnote{Here $r$ is not count because it is the holographic direction: the coordinate of $r$ is relate to the energy scale of the dual field theory, in the spirit of gauge/gravity correspondence.} for the boundary fluid. So we can not dimensionally reduce AdS$_3$, one should begin at least with AdS$_4$. And also, the number of $q$ can not be larger than $p-1$, by the same token.

\section{The compactified AdS black hole}

The full action of the AdS$_{p+2}$ black hole contains the bulk action, the Gibbons-Hawking term and the counter term. It can be written as
\begin{align}\label{eq: AdS(p+2) full action}
	S = &\; \frac1{2\kappa_{p+2}^2} \int d^{p+2}x \sqrt{-G} \left[\mathcal R+\frac{p(p+1)}{L^2}\right]- \frac{1}{\kappa_{p+2}^2} \int d^{p+1}x \sqrt{-H} \mathcal K  \cr
	    & + \frac{1}{\kappa_{p+2}^2} \int d^{p+1}x \sqrt{-H}\, \frac{p}{L}.
\end{align}
Here $G_{\hat M \hat N}$ and $\mathcal R$ are the metric tensors and the Ricci scalar of the AdS$_{p+2}$ black hole. Its boundary is defined on a $p+1$ dimensional hyperplane with $H_{\hat M \hat N}$ and $\mathcal K$ are the induced metric and extrinsic curvature, respectively. Compared with \cite{Chamblin9903}, the Gibbons-Hawking term and the counter term can be seen as the action of a domain wall placed at the boundary. The difference of this work from \cite{Chamblin9903} is that the domain wall in \cite{Chamblin9903} is placed inside the bulk and it splits the background into two parts. The action of the domain wall contains the extrinsic curvature on both sides. But here in \cref{eq: AdS(p+2) full action}, since the domain wall is placed near the boundary, we only count the extrinsic curvature on one side---the outer side. It is natural to only consider the inner side because one should not care about what happens outside the boundary. But in practice, we are accustomed to normal vectors pointing to the positive $r$ direction. So here we only count the outside part.

The solution of the bulk action in \cref{eq: AdS(p+2) full action} is the metric of AdS$_{p+2}$ black hole:
\begin{align} \label{eq: AdS(p+2) BH metric}
  ds^2 = \frac{r^2}{L^2}\left( -f(r) dt^2 + \delta_{ij}dx^idx^j \right) + \frac{L^2}{r^2} \frac{dr^2}{f(r)} + \frac{r^2}{L^2} \delta_{mn} dy^m dy^n,
\end{align}
where $f(r) = 1 - \frac{r_H^{p+1}}{r^{p+1}}$. The indices $i,\; j$ run from 1 to $p-q$ and $m,\; n$ from 1 to $q$. We compactified $q$ of the $p+1$ spatial directions as a $q$ dimensional torus $\mathbf T^q$. Such that after integrating $\mathbf T^q$, the above line element will become $\text{CR}_{p-q+2}$, i.e. $\text{AdS}_{p+2} \to  \text{CR}_{p-q+2} \times \mathbf T^q$. The coordinate system is set as $x^{\hat M} = (x^M,\ y^m)$, of which $x^M$ is the coordinate of $\text{CR}_{p-q+2}$ and $y^m$ of $\mathbf T^q$.

In order to reduce the metric of AdS$_{p+2}$ black hole to a $(p-q+2)$-dimensional spacetime, we use the following reduction ansatz:
\begin{align}\label{eq: (p+2)D reduction ansatz}
	ds^2 = e^{2\alpha_1 A} g_{MN} dx^M dx^N + e^{2\alpha_2 A} \delta_{mn} dy^m dy^n,
\end{align}
and
\begin{align}\label{eq: (p+1)D reduction ansatz}
	ds^2 = e^{2\alpha_1 A} h_{MN} dx^M dx^N + e^{2\alpha_2 A} \delta_{mn} dy^m dy^n,
\end{align}
to separately reduce the bulk and boundary sectors of the $(p+2)$-dimensional theory. So the quantities of $(p+2)$-dimensional AdS black hole can be reduced as
\begin{align}
  \sqrt{-G} &= e^{[(p-q+2) \alpha_1 + q \alpha_2] A} \sqrt{-g}, \qquad \sqrt{-H} = e^{[(p-q+1) \alpha_1 + q \alpha_2] A} \sqrt{-h}, \cr
   \mathbf n_{\hat M} &= (\mathbf n_M, \mathbf n_m) = \left( e^{\alpha_1 A} n_M, 0 \right), \qquad \mathcal K = e^{- \alpha_1 A} K.
\end{align}
Here $\mathbf n_{\hat M} = \frac{\nabla_{\hat M}r}{\sqrt{ G^{\hat N\hat P} \nabla_{\hat N}r \nabla_{\hat P}r }}$ is the unit norm on the $(p+1)$-dimensional boundary pointing outwards and $n_M$ is the unit norm of the reduced theory.

Then \cref{eq: AdS(p+2) full action} becomes
\begin{align}
  S =&\, \frac{V_q}{2\kappa_{p+2}^2} \int d^{p-q+2}x \sqrt{-g} e^{[(p-q) \alpha_1 + q \alpha_2] A} \bigg[ R - 2 \big( (p-q+1) \alpha_1 + q \alpha_2 \big) \nabla^2 A  \cr
  & - \Big( (p-q+1)(p-q) \alpha_1^2 + 2q(p-q) \alpha_1\alpha_2 + q(q+1) \alpha_2^2 \Big) (\partial A)^2 + \frac{p(p+1)}{L^2} e^{2\alpha_1 A} \bigg] \cr
  & - \frac{V_q}{\kappa_{p+2}^2} \int d^{p-q+1}x \, e^{[(p-q) \alpha_1 + q \alpha_2] A} \sqrt{-h} \, e^{- \alpha_1 A} K \cr
  & + \frac{V_q}{\kappa_{p+2}^2} \int d^{p-q+1}x \, e^{[(p-q) \alpha_1 + q \alpha_2] A} \sqrt{-h} \, \frac{p}{L}.
\end{align}
In Einstein frame, the Ricci scalar should not couple with scalar field directly, which requires that we should set $(p-q) \alpha_1 + q \alpha_2 = 0$. Thus one may choose
\begin{align}\label{eq: alpha(1,2) value}
	\alpha_1 = - \frac{q}{p-q}, \qquad \alpha_2 = 1.
\end{align}
Then one can get the reduced action in $p-q+2$ dimensions as
\begin{align}\label{eq: (p-q+2) dimensional reduced action}
  S =&\, \frac1{2\kappa_{p-q+2}^2} \int d^{p-q+2}x \sqrt{-g} \left[ R + \frac{2q}{p-q} \nabla^2 A - \frac{pq}{p-q} (\partial A)^2 + \frac{p(p+1)}{L^2} e^{-\frac{2q}{p-q}A} \right] \cr
	    & - \frac1{\kappa_{p-q+2}^2} \int d^{p-q+1}x \sqrt{-h} \, K + \frac1{\kappa_{p-q+2}^2} \int d^{p-q+1}x \sqrt{-h} \, \frac{p}{L} \, e^{-\frac{q}{p-q}A}.
\end{align}
Here
\begin{align}
  \frac1{2\kappa_{p-q+2}^2} = \frac{V_q}{2\kappa_{p+2}^2}
\end{align}
is the surface gravity in $p-q+2$ dimensions. The background scalar $A$ comes from the reduction process. It represents the radius of the compact part in AdS$_{p+2}$ black hole metric \cref{eq: AdS(p+2) BH metric}, i.e. $\mathbf T^q$. By setting $q=0$ in \cref{eq: (p-q+2) dimensional reduced action}, one will get the action of AdS$_{p+2}$ black hole \cref{eq: AdS(p+2) full action} again.

Note $\nabla^2 A$ in \cref{eq: (p-q+2) dimensional reduced action} is a boundary term and can be dropped, because the domain wall here is at the boundary. But when the domain wall is placed in the middle of the bulk spacetime like in \cite{Chamblin9903}, it can not be ignored and will also have a corresponding term on the domain wall as
\begin{align}
  \frac1{\kappa_{p-q+2}^2} \int d^{p-q+1}x \sqrt{-h} \, \frac{q}{p-q} n^\mu \nabla_\mu A.
\end{align}
The action \cref{eq: (p-q+2) dimensional reduced action} describes a theory of Einstein gravity coupled with a background scalar and a dynamical domain wall sitting at the boundary. Thus the $(p-q+2)$-dimensional reduced theory (or the compactified AdS black hole) is really the Chamblin-Reall model, which is denoted as CR$_{p-q+2}$.
The Gibbons-Hawking term and the counter term in \cref{eq: (p-q+2) dimensional reduced action} can now be separately understood as the extrinsic curvature and the scalar potential of the dynamical domain wall. The counter term may also be seen as the interaction between the background scalar and the domain wall.

From \cref{eq: (p-q+2) dimensional reduced action} one can derive the EOMs for $g_{MN}$ and $A$ in the reduced theory as
\begin{align}
    E_{MN} &- T_{MN} = 0, \label{eq: EOM Einstein eq} \\
    \nabla^2 A &- \frac{p+1}{L^2} e^{-\frac{2q}{p-q}A}=0. \label{eq: EOM A}
\end{align}
Here $T_{MN}$ is the bulk energy-momentum tensor in CR$_{p-q+2}$, whose explicit form is
\begin{align}
  T_{MN} =&\frac{pq}{p-q} \left(\partial_M A \partial_N A - \frac12 g_{MN} (\partial A)^2 \right) + \frac{p(p+1)}{2L^2} g_{MN} e^{-\frac{2q}{p-q}A}.
\end{align}
By comparing \cref{eq: AdS(p+2) BH metric} with \cref{eq: (p+2)D reduction ansatz}, one can get the classical background of $(p-q+2)$-dimensional compactified AdS black hole, which is also the background of CR$_{p-q+2}$, as:
\begin{align}
	ds^2 &= \left( \frac r{L} \right)^{\frac{2p}{p-q}} \left( -f(r) dt^2 + d \vv x^2 \right) + \left( \frac{L}{r} \right)^\frac{2p-4q}{p-q} \frac{dr^2}{f(r)},  \label{eq: (p-q+2) dimensional reduced metric of cAdS BH} \\
	e^A &= \frac{r}{L}. \label{eq: A of the reduced background of cAdS BH}
\end{align}
It can be checked that the above background solutions \cref{eq: (p-q+2) dimensional reduced metric of cAdS BH,eq: A of the reduced background of cAdS BH} solve \cref{eq: EOM Einstein eq,eq: EOM A}. Also note that the background of CR$_{p-q+2}$ has a non-trivial scalar field $A$, suggesting the non-conformality. \cref{eq: (p-q+2) dimensional reduced metric of cAdS BH} is no longer asymptotically AdS, which can be seen from the power of the factor $r/L$ in the metric. Note when setting $q=0$ in \cref{eq: (p-q+2) dimensional reduced metric of cAdS BH}, one gets the metric of AdS$_{p-q+2}$ black hole. It is $q$ that makes the metric \cref{eq: (p-q+2) dimensional reduced metric of cAdS BH} deviate from asymptotically AdS.

The metric of compactified AdS black hole \cref{eq: (p-q+2) dimensional reduced metric of cAdS BH} is valid for all $p \geq 2$ and $1 \leq q \leq p-1$ and it is asymptotically flat for all the valid $p$ and $q$. This can be checked by calculating the Ricci and Kretschmann scalars for \cref{eq: (p-q+2) dimensional reduced metric of cAdS BH}:
\begin{align}
  R =& - \frac{p}{p-q} \frac{(p+2) (p-q+1) r^{p+1} + q r_H^{p+1}}{r^{p+1} L^2} \left( \frac{L}{r} \right)^\frac{2q}{p-q}, \\
  R_{MNPQ} R^{MNPQ} =&\; \frac{p^2}{(p-q)^3} \bigg[ 2(p^2+2p-2q) (p-q+1) r^{2p+2} + 4q (p-2q) r^{p+1} r_H^{p+1} \cr
  & + \Big( (p-q)^2 (p^3-p^2q-2pq) - (p-q) (p^2-2pq-q^2) + 2q^2 \Big) r_H^{2p+2} \bigg] \cr
  & \times \frac{1}{r^{2p+2} L^4} \left( \frac Lr \right)^\frac{4q}{p-q}.
\end{align}
For the given range of $p$ and $q$, these two scalars always approach zero as $r\to \infty$, meaning \cref{eq: (p-q+2) dimensional reduced metric of cAdS BH} is indeed asymptotically flat.

To compactify the AdS black hole and to deform it by adding background scalar field(s) are two ways to get non-conformal backgrounds from the AdS black hole. Each way has its own advantages and limitations. For the compactifying way, the background will remain analytic and will produce exact and analytic results in full non-conformal forms. The calculation is usually direct at every step. Analytical results are convenient to check the identities among the second-order transport coefficients but they are often far from the case in the real world.
For the deforming way, one needs to search for numerical solutions and the results are all in numerical forms. The calculation is not very direct since the functions that appear in the calculating process are usually unknown. Numerical results are not convenient for checking the identities but they are usually closer to the real cases, at least qualitatively, than the analytical results.

Toroidally compactifying the AdS black hole is a simple way to generalize the AdS black hole both analytically and non-conformally. Compared with the metric of compactified Dp-brane \cite{Wu1604,Wu2012}, the AdS black hole metric in planar limit does not have the spherical part. This makes the process of dimensional reduction for AdS black holes much easier than that of compactified Dp-brane.

The toroidal compactification that we use for both the D-brane and AdS black hole is special in that the reduction ansatzes of these two cases are diagonal. More general way of compactifying, e.g. adding non-diagonal elements in reduction ansatz will lead to the appearance of additional background fields in the reduced bulk gravity, whose roles in the boundary relativistic fluids may not be clear. Since the reduced gravity theory for both the compactified Dp-brane and AdS black hole is Einstein gravity coupled with one background scalar. The scalar field brings the theory into a non-conformal regime and the metric tensor gives the boundary stress-energy tensor in the sense of the Brown-York tensor.

We are of the opinion that the compactified AdS black hole will be very useful in other subjects of holographic studies.

\section{Connections between the non-conformal backgrounds and the Chamblin-Reall model}

In this section, we will uncover the truth that the compactified AdS black hole, the reduced compactified Dp-brane and all the other non-conformal supergravity backgrounds that are used in the non-conformal holographic studies of the relativistic hydrodynamics are all Chamblin-Reall type backgrounds.

\subsection{The compactified AdS black hole}

The compactified AdS black hole can be understood as a classical Einstein-scalar-domain wall gravitational system with the domain wall placed as the boundary. The action and background can be written as
\begin{align}
  S =&\, \frac1{2\kappa_{p-q+2}^2} \int d^{p-q+2}x \sqrt{-g} \left[ R - \frac{pq}{p-q} (\partial A)^2 + \frac{p(p+1)}{L^2} e^{-\frac{2q}{p-q}A} \right] \cr
	& - \frac1{\kappa_{p-q+2}^2} \int d^{p-q+1}x \sqrt{-h} \, \left( K - \frac{p}{L} \, e^{-\frac{q}{p-q}A} \right), \cr
  ds^2 =& \left( \frac r{L} \right)^{\frac{2p}{p-q}} \left( -f(r) dt^2 + d \vv x^2 \right) + \left( \frac{L}{r} \right)^\frac{2p-4q}{p-q} \frac{dr^2}{f(r)}, \qquad 	e^A = \frac{r}{L}.
\end{align}
The second term in the above action is the domain wall action. Note the exponent in the coupling term between the domain wall and the scalar is $- \frac{q}{p-q} A$, which is half of the exponent in the scalar potential.

Now we rescale the scalar field by $\sqrt{pq \over p-q} A = \frac{1}{\sqrt2} \phi$, then the compactified AdS black hole system becomes
\begin{align}\label{eq: compactified AdS BH in CR form}
  S =&\, \frac1{2\kappa_{p-q+2}^2} \int d^{p-q+2}x \sqrt{-g} \left[ R - \frac12 (\partial \phi)^2 + \frac{p(p+1)}{L^2} e^{- \gamma \phi} \right] \cr
	& - \frac1{\kappa_{p-q+2}^2} \int d^{p-q+1}x \sqrt{-h} \, \left( K - \frac{p}{L} \, e^{- \frac{\gamma}{2} \phi} \right), \cr
  ds^2 =& \left( \frac r{L} \right)^{\frac{2p}{p-q}} \left( -f(r) dt^2 + d \vv x^2 \right) + \left( \frac{L}{r} \right)^\frac{2p-4q}{p-q} \frac{dr^2}{f(r)}, \qquad 	e^\phi = \left( \frac{r}{L} \right)^{p \gamma}.
\end{align}
Here
\begin{align}\label{eq: gamma^2 in CR form of comp. AdS BH}
  \gamma^2 = \frac{2q}{p(p-q)} = \frac{2}{p-q} \left( 1- (p-q) c_s^2 \right) = \frac{2}{p-q}\delta_\text{nonconf.},
\end{align}
where $\delta_\text{nonconf.} = 1- (p-q) c_s^2$ is the parameter of non-conformal corrections used in \cite{Bigazzi1006,Kleinert1610} and $\gamma$ is the parameter used in \cite{Gubser0804PRD,Gubser0806,Bigazzi1006,Kleinert1610}. If substituting the sound speed for the compactified AdS black hole as $c_s^2 = \frac{1}{p}$, one will have $\delta_\text{nonconf.} = \frac qp$. In the special case that the bulk spacetime is CR$_5$, i.e. $q=p-3$, we have $\delta_\text{nonconf.} = 1 - \frac3p$. We can generalize the second-order transport coefficients for CR$_5$ produced in \cite{Kleinert1610} into general dimensions by using  $\delta_\text{nonconf.} = \frac qp$ instead of $\delta_\text{nonconf.} = 1 - \frac3p$ in a later section.

In the present form, \cref{eq: compactified AdS BH in CR form} is the $(p-q+2)$-dimensional Chamblin-Reall model reduced from the AdS black hole, which we may denote as CR$_{p-q+2}$. When setting $q=p-3$ and $p=D-1$, one can reproduce the background of CR$_5$ reduced from the AdS$_{D+1}$ black hole that is used in \cite{Gubser0804PRD,Gubser0806,Bigazzi1006,Kleinert1610}.

\subsection{The compactified Dp-brane}

The $(p-q+2)$-dimensional reduced compactified near-extremal Dp-brane action can be written as \cite{Wu2012}
\begin{align}\label{eq: compactified Dp-brane action}
  S =&\; \frac1{2\kappa_{p-q+2}^2} \int d^{p-q+2}x \sqrt{-g} \left[ R - \frac{8(8-p+q)}{p-q} (\partial A)^2 - \frac{(8-p)(8-p+q)}{q} (\partial B)^2 \right. \cr
  &\left. - \frac12 (\partial \phi)^2 - V(\phi,A,B) \right] - \frac1{\kappa_{p-q+2}^2} \int d^{p-q+1}x \sqrt{-h} \left( K - \frac{9-p}{2L_p} e^{- \frac{8-p+q}{p-q} A - \frac{p-3}{4(7-p)} \phi} \right),
\end{align}
with the scalar potential
\begin{align}
  V(\phi,A,B) = \frac{(7-p)^2}{2L_p^2} e^{\frac{p-3}{2} \phi - \frac{2(p-q)(7-p) + 16}{p-q}A - 2(8-p) B} - \frac{(7-p)(8-p)}{L_p^2} e^{- \frac{16}{p-q}A - 2B}.
\end{align}
The background solution of \cref{eq: compactified Dp-brane action} is
\begin{align}
  ds^2 &= \left( \frac r{L_p} \right)^{\frac{9-p}{p-q}} \left( -f(r) dt^2 + d \vv x^2 \right) + \left( \frac{r}{L_p} \right)^\frac{(p^2-8p+9) + q(7-p)}{p-q} \frac{dr^2}{f(r)},  \label{eq: (p-q+2) dimensional reduced metric} \\
   e^{\phi} &= \left( \frac r{L_p} \right)^\frac{(p-3)(7-p)}{4}, \qquad e^{A} = \left( \frac r{L_p} \right)^{\frac{(p-3)^2}{16} + \frac{q(5-p)}{2(8-p+q)}}, \qquad e^B = \left( \frac r{L_p} \right)^{- \frac{q(5-p)}{2(8-p+q)}}.
\end{align}
The three background scalars are not independent of each other by
\begin{align}\label{eq: relations of A and B with dilaton}
  A &= \left[ \frac{p-3}{4(7-p)} + \frac{2q(5-p)}{(8-p+q)(p-3)(7-p)} \right] \phi, \cr
  B &= \frac{- 2q(5-p)}{(8-p+q)(p-3)(7-p)} \phi.
\end{align}
After substituting the above into the action and background of compactified Dp-brane, one has
\begin{align}
  S =&\; \frac1{2\kappa_{p-q+2}^2} \int d^{p-q+2}x \sqrt{-g} \bigg[ R - \frac{4(9-p) ( (p-3)^2 + q(5-p) )}{(p-q) (p-3)^2 (7-p)^2} (\partial \phi)^2  \cr
  & + \frac{(7-p)(9-p)}{2L_p^2} e^{- \frac{4(p-3)^2 + 4q(5-p)}{(p-q) (p-3) (7-p)} \phi} \bigg] \cr
  &- \frac1{\kappa_{p-q+2}^2} \int d^{p-q+1}x \sqrt{-h} \left( K - \frac{9-p}{2L_p} e^{- \frac{2(p-3)^2 + 2q(5-p)}{(p-q) (p-3) (7-p)} \phi} \right).
\end{align}

In order to connect to the Chamblin-Reall model, one still needs to rescale the scalar as
\begin{align}
  \sqrt{\frac{4(9-p) ( (p-3)^2 + q(5-p) )}{(p-q) (p-3)^2 (7-p)^2}} \phi = \frac{1}{\sqrt2} \varphi.
\end{align}
Thus the action finally becomes the Chamblin-Reall type
\begin{align}\label{eq: compactified Dp-brane action in CR form}
  S =&\; \frac1{2\kappa_{p-q+2}^2} \int d^{p-q+2}x \sqrt{-g} \bigg[ R - \frac12 (\partial \varphi)^2 + \frac{(7-p)(9-p)}{2L_p^2} e^{- \gamma \varphi} \bigg] \cr
  & - \frac1{\kappa_{p-q+2}^2} \int d^{p-q+1}x \sqrt{-h} \left( K - \frac{9-p}{2L_p} e^{- \frac{\gamma}{2} \varphi} \right).
\end{align}
The background now only contains one scalar field
\begin{align}\label{eq: compactified Dp-brane background in CR form}
  ds^2 &= \left( \frac r{L_p} \right)^{\frac{9-p}{p-q}} \left( -f(r) dt^2 + d \vv x^2 \right) + \left( \frac{r}{L_p} \right)^\frac{(p^2-8p+9) + q(7-p)}{p-q} \frac{dr^2}{f(r)},  \\
  e^{\varphi} &= \left( \frac r{L_p} \right)^{\frac{9-p}{2} \gamma}.
\end{align}
Here the parameter $\gamma$ is derived as
\begin{align}
  \gamma^2 = \frac{2(p-3)^2 + 2q(5-p)}{(p-q) (9-p)}.
\end{align}
It is interesting to note that $\gamma^2$ is the numerical part of the bulk viscosity $\zeta$ of the compactified Dp-brane, as can be checked by referring to \cite{Wu2012}. This will also be true for the compactified AdS black hole, as one can check by using the results of first-order transport coefficients offered in the next section.

By the present form, the action and background solution shown in \cref{eq: compactified Dp-brane action in CR form,eq: compactified Dp-brane background in CR form} is another $(p-q+2)$-dimensional Chamblin-Reall model with different origins.

\subsection{The other non-conformal backgrounds}

In this subsection, we will show that the backgrounds that are used in the early non-conformal studies of strongly coupled relativistic fluids are all Chamblin-Reall type backgrounds, but with more than one independent background scalar field.

\subsubsection{The $\mathcal N=2^*$ model}

The 5-dimensional reduced action that is dual to $\mathcal N=2^*$ supersymmetric gauge theory can be written as
\cite{Buchel0311,Buchel0406200,Benincasa0507,Buchel0812,Buchel0908,Buchel1110,Buchel1503}
\begin{align}\label{eq: N=2^* action}
  S =&\; \frac{1}{2\kappa_5^2} \int d^5x \sqrt{-g} \Big( R - 12(\partial \phi)^2 - 4(\partial \chi)^2 - V(\phi,\chi) \Big) \cr
  & - \frac1{\kappa_5^2} \int d^5x \sqrt{-h} \Big( K - P(R_4[h],\phi,\chi) \Big).
\end{align}
Here $R_4[h]$ is the Ricci scalar on the domain wall and $\phi,\; \chi$ are the two background scalars. $P(R[h],\phi,\chi)$ is a polynomial of $R_4[h]$, $\phi$ and $\chi$ whose explicit form can be found in \cite{Buchel0406200}. The potential of the background scalars is \cite{Buchel1503}
\begin{align}
  V(\phi,\chi) = \frac{1}{L^2} \left( e^{8\phi} \sinh^2 2\chi - 8 e^{2\phi} \cosh 2\chi - 4 e^{-4\phi} \right).
\end{align}

By the definition of the hyperbolic function, one can see that this potential is also a Liouville potential. The only difference between the cases of compactified AdS black hole and the compactified Dp-brane is that now the scalar potential contains two independent scalars and more than one exponential term.
The conformal limit is achieved by setting both of the two scalars to 0, leaving us $V = - \frac{12}{L^2}$, which is the potential of the AdS$_5$ black hole. So the gravitational background that solves the EOMs of \cref{eq: N=2^* action} should be asymptotically AdS$_5$, which may be written like
\begin{align}
  ds^2 = e^{2A(z)} \frac{L^2}{z^2} \left( - f(z) dt^2 + d \vv x^2 + \frac{dz^2}{f(z)} \right).
\end{align}
Here $f(z)$ is the blackening factor and $A(z)$ measures the extent of deviation from the AdS$_5$ black hole. One needs to solve 4 unknown functions: $A(z),\ f(z),\ \phi(z)$, and $\chi(z)$ in this background. Generally, there is no exact and analytic solution for this model, most of the works done analytically are also perturbatively as in
\cite{Buchel0311,Buchel0406200,Benincasa0507,Buchel0812,Buchel0908,Buchel1110}.

By the statement above, one can see that the 5-dimensional reduced bulk theory belongs to the Chamblin-Reall model.

Before we move to the next two non-conformal supergravity models, we would like to mention another background with also two background scalar fields. The interesting point is that it is not non-conformal, but conformal. This model is the near D3-brane with D-instanton smeared on its world-volume, sometimes it is called the D(-1)-D3 or D-instanton-D3 model. It is originally 10-dimensional, but after reducing it to 5 dimensions, one has the action with the classical solution as
\begin{align}\label{eq: D(-1)D(3) 5D reduced action}
  S = \frac1{2\kappa_{5}^2} \int d^{5}x \sqrt{-g} \left[R- \frac{1}{2}(\partial \phi)^2 +\frac12 e^{2\phi}(\partial \chi )^2 + \frac{12}{L_3^2}\right] - \frac{1}{\kappa_{5}^2} \int d^{4}x \sqrt{-h} \left( K - \frac{3}{L_3} \right)
\end{align}
and
\begin{align}\label{eq: D(-1)D(3) 5D reduced background}
	ds^2 = \frac{r^2}{L_3^2} \left( -f(r) dt^2 + d \vv x^2 \right) + \frac{L_3^2}{r^2} \frac{dr^2}{f(r)}, \quad
    e^\phi  = H_{-1}, \quad \chi = 1 - e^{-\phi}.
\end{align}
From the fact that the scalar potential in the bulk action is $V = - \frac{12}{L_3^2}$, the same as the AdS$_5$ black hole. We can see the background should be conformal, though it has two dependent non-trivial background scalars $\phi$ and $\chi$. The counter term in the above is also a constant: suggesting the conformality of the background, too. The action \cref{eq: D(-1)D(3) 5D reduced action} is not Chamblin-Reall type for the scalar potential is constant, not Liouville potential.

But why does the non-trivial dilaton and axion do not bring the background into a non-conformal regime? The reason lies in the form of $\chi$, by $\chi=1-e^{-\phi}$, one has $\partial \chi = e^{-\phi} \partial\phi$, which leads to $(\partial\phi)^2 = e^{2\phi} (\partial \chi)^2$. So in the on-shell action of D(-1)-D3 model, the profile of the axion exactly cancels the dilaton's, making the on-shell action becomes
\begin{align}\label{eq: D(-1)D(3) 5D reduced action onshell}
  S = \frac1{2\kappa_{5}^2} \int d^{5}x \sqrt{-g} \left[R + \frac{12}{L_3^2}\right] - \frac{1}{\kappa_{5}^2} \int d^{4}x \sqrt{-h} \left( K - \frac{3}{L_3} \right),
\end{align}
which is exactly the same as the AdS$_5$ black hole. Thus the background is conformal.

The physical account for the fact that the addition of smeared D(-1)-branes does not make the background non-conformal may be drawn like this. The D(-1)-brane, which carries a global ``electric" charge in string theory \cite{Gibbons9511}, is the instanton solution in string theory. Its magnetic dual is the D7-brane, which is the cosmic string solution in string theory \cite{Greene1990NPB}. Both the instanton and cosmic strings are topological objects. But non-conformality is a local effect, it should not be affected by the presence of topological objects.

\subsubsection{The D3-D7 model}

The D3-D7 model is originally 10-dimensional, it is type IIB supergravity sourced by $N_3$ D3-branes with the correction of $N_7$ D7-branes' back reaction. After reducing on the compact manifold $\mathbf{T}^1 \times {\rm CP}^2$ \cite{Bigazzi0909,Bigazzi0912}, one gets a 5-dimensional Einstein gravity coupled with 3 background scalars. The action is
\begin{align}\label{eq: D3-D7 5D reduced action}
  S =&\; \frac{1}{2\kappa_5^2} \int d^5x \sqrt{-g} \left( R - \frac12 (\partial\phi)^2 - \frac{40}{3} (\partial A)^2 - 20 (\partial B)^2 - V(\phi,A,B) \right) \cr
  & - \frac{1}{\kappa_{5}^2} \int d^{4}x \sqrt{-h} \bigg[ K - \bigg( Q_3 e^{\frac{20}{3} A} + Q_7 e^{\phi + \frac83 A - 4B} - 4 e^{\frac83 A + 6B} - 6 e^{\frac83 A - 4B} \cr
  & - \frac12 C R_4[h] \bigg) \bigg].
\end{align}
Here $C \approx 1 + \frac{23}{108} \epsilon_* - \frac{371}{23328} \epsilon_*^2 + \mathcal O\left( \frac{1}{r^4} \right)$. $\epsilon_*$ is an expansion parameter at some specially chosen UV cutoff of the model, the details can be found in \cite{Bigazzi0912}. The Liouville potential in the above action has the form
\begin{align}\label{eq: D3-D7 5D scalar potential}
  V(\phi,A,B) = \frac{1}{L^2} \left( \frac{Q_7^2}{2} e^{2\phi + \frac{16}{3} A - 8B} + 4 Q_7
   e^{\phi + \frac{16}{3}A + 2B} + 8 e^{\frac{40}{3} A } + 4 e^{\frac{16}{3} A + 12 B} - 24 e^{\frac{16}{3}A + 2B} \right).
\end{align}
Here $Q_7 = \frac{g_s}{2\pi} N_7$. By the form of \cref{eq: D3-D7 5D reduced action,eq: D3-D7 5D scalar potential}, the 5-dimensional reduced theory of D3-D7 model is Chamblin-Reall type with three background scalars. The background metric and scalars are solved in terms of series of $\epsilon_h$ to the second order:
\begin{align}
  ds^2 &= e^{- \frac{10}{3} A} \frac{r^2}{L^2} ( -f(r) dt^2 + d \vv x^2) + e^{- \frac{40}{3} A} \frac{L^2 dr^2}{r^2 f(r)}, \qquad f(r) = 1 - \frac{r_H^4}{r^4}, \\
  \phi &= - \epsilon_h \ln\frac{r_H}{r} - \epsilon_h^2 \left[ \frac16 \ln\frac{r_H}{r} - \frac12 \left( \ln\frac{r_H}{r} \right)^2 - \frac{1}{16} \operatorname{Li_2}\left( 1-\frac{r_H^4}{r^4} \right) \right], \\
  A &= - \frac{1}{40} \epsilon_h - \frac{1}{120} \left( 1 - 3 \ln\frac{r_H}{r} \right) \epsilon_h^2, \\
  B &= - \frac{1}{60} \epsilon_h + \frac{1}{720} \left( 1 + 12 \ln\frac{r_H}{r} \right) \epsilon_h^2.
\end{align}
The $\epsilon_h$ here is the same expansion parameter of the model but now taken the value at the horizon \cite{Bigazzi0912}. From the above, we can see that the 3 background scalars in the 5-dimensional reduced form of the D3-D7 model are independent of each other.

The numerical value of the kinetic terms of the scalars in \cref{eq: D3-D7 5D reduced action} are completely the same as that of D(4-1)-brane \cite{Wu1604}. The compact manifold of D(4-1)-brane is $ \mathbf T^1 \times \mathbf  S^4$, which is different from that of the D3-D7 model. But the numbers of dimensions of the compact submanifolds of these two models are the same. Thus one may draw a conclusion that the kinetic terms of the scalars from dimensional reduction are determined only by the dimensions of the compact submanifold, not by the details of the geometry. For the D(4-1)-brane and the D3-D7 model, the dimensions of the compact submanifolds are both $(1+4)$, although the geometries are different: one is $\mathbf S^4$, the other is ${\rm CP}^2$.

\subsubsection{The Klebanov-Tseytlin background}

The last model is the Klebanov-Tseytlin background. It is the gravity dual of the cascading gauge theory whose 5-dimensional reduced bulk action is
\begin{align}\label{eq: Klebanov-Tseytlin model 5D reduced bulk action}
  S =&\; \frac{1}{2\kappa_5^2} \int d^5x \sqrt{-g} \bigg( R - \frac12 (\partial\phi)^2 - \frac12 e^{-\phi + 4A +4B} (\partial \chi)^2 - \frac{40}{3} (\partial A)^2 - 20 (\partial B)^2 \cr
  & - V(\phi,\chi,A,B) \bigg),
\end{align}
where the scalar potential reads
\begin{align}
  V(\phi,\chi,A,B) =&\; \frac{1}{L^2} \Big[ Q^2 \left( e^{\phi + \frac{28}{3} A - 4B} + e^{\frac{40}{3} A + 2\chi} \right) + 4\sqrt2 Q e^{\frac{40}{3} A + \chi} + 8 e^{\frac{40}{3} A} \cr
  & + 4 e^{\frac{16}{3} A + 12B} - 24 e^{\frac{16}{3} A + 2B} \Big].
\end{align}
This model has four background scalar fields. The Gibbons-Hawking term and counter term of this model are extremely complicated for us to record here. The readers who are interested can find it in \cite{Aharony0506}. When $Q=0$, all the scalars are 0, leaving the potential $V = - \frac{12}{L^2}$, which is still the same as that of the AdS$_5$ black hole. Complicate as \cref{eq: Klebanov-Tseytlin model 5D reduced bulk action} is, the 5-dimensional reduced theory of Klebanov-Tseytlin background is still a Chamblin-Reall type model, with the number of independent background scalars reaches four.

\section{The first order results}

Now we are back on the discussions about compactified AdS black hole and will calculate its transport coefficients. This section will deal with the first order. The global on-shell metric written in Eddington-Finkelstein coordinate in boosted form is
\begin{align}\label{eq: global on-shell metric}
    ds^2 =& -r^\frac{2p}{p-q} [ f(r_H(x),r) + k(r_H(x), u^\alpha(x), r) ] u_\mu(x) u_\nu(x) dx^\mu dx^\nu \cr
    & - 2 r^\frac{2p}{p-q} P_\mu^\rho(u^\alpha(x)) w_\rho(u^\alpha(x), r) u_\nu(x) dx^\mu dx^\nu  \cr
    & + r^\frac{2p}{p-q} [ P_{\mu\nu}(u^\alpha(x)) + \alpha _{\mu\nu}(r_H(x), u^\alpha(x), r) + h(r_H(x), u^\alpha(x), r) P_{\mu\nu}(u^\sigma(x)) ] dx^\mu dx^\nu \cr
    & - 2 r^\frac{2q}{p-q} [ 1 - j(r_H(x), u^\alpha(x), r) ] u_\mu(x) dx^\mu dr.
\end{align}
Here the perturbation ansatz $k,\ j$, and $w_\rho$ have an extra minus sign compared with \cite{Wu1807,Wu2012}. In the above,
\begin{align}
  u^\mu = \frac{(1, \vv \beta)}{\sqrt{1-\vv \beta^2}}, \qquad P_{\mu\nu} = \eta_{\mu\nu} + u_\mu u_\nu.
\end{align}

Then put the first order expanded on-shell metric
\begin{align}\label{eq: 1st order expanded metric}
  ds^2 =&\; r^\frac{2p}{p-q} \bigg[ - \bigg( f(r) - \frac{(p+1)r_H^{p}}{r^{p+1}}\delta r_H + k^{(1)}(r) \bigg) dv^2 \cr
  & + 2 \big( (f-1) \delta \beta_i + w^{(1)}_i(r) \big) dvdx^i  + (\delta_{ij} + \alpha_{ij}^{(1)}(r) + h^{(1)}(r) \delta_{ij}) dx^idx^j \bigg] \cr
  & + 2 r^\frac{2q}{p-q} (1 - j^{(1)}(r)) dvdr - 2 r^\frac{2q}{p-q} \delta\beta_i dx^idr
\end{align}
into the traceless tensor components of Einstein equation
\begin{align}\label{eq: Einstein eq(ij)}
  E_{ij} - \frac1{p-q} \delta_{ij} \delta^{kl} E_{kl} - \left( T_{ij} - \frac1{p-q} \delta_{ij} \delta^{kl} T_{kl} \right) = 0,
\end{align}
we will have
\begin{align}\label{eq: diff eq of 1st order alpha(ij)}
  \partial_r (r^{p+2} f(r) \partial_r \alpha^{(1)}_{ij}(r)) + 2p r^\frac{p-1}2 \sigma_{ij} = 0.
\end{align}
The above equation of the tensor perturbation does not contain $q$, which should not be a surprise as it has been pointed out in \cite{Wu2012} that dimensional reduction does not affect the tensor and vector part of the perturbations. Thus the solutions for \cref{eq: diff eq of 1st order alpha(ij)} are the same as the AdS$_{p+2}$ black hole, which can be found in \cite{Haack0806}. By the SO$(p-q)$ invariance, one can set $\alpha^{(1)}_{ij}(r) = F(r) \sigma_{ij}$, then the first order tensor perturbation can be solved as
\begin{align} \label{eq: F(r) solutions}
  F(r) =&\; \frac1{3r_H} \left( 2\sqrt3 \arctan \frac{ \sqrt{3} r_H }{2r + r_H}+ 3 \ln \frac{r^2 + r r_H + r_H^2}{r^2} \right), \quad \text{(p=2)} \cr
  F(r) =&\; \frac1{2r_H} \left[ 2\arctan \frac{r_H}{r} + \ln\frac{(r + r_H)^2 (r^2 + r_H^2)}{r^4} \right], \quad \text{(p=3)} \cr
  F(r) =&\; \frac{1}{5r_H} \bigg[ 4\sin \frac{2\pi}{5} \arctan \frac{r_H \sin \frac{2\pi}{5}}{r - r_H \cos \frac{2\pi}{5}} + 4\sin \frac{\pi}{5} \arctan \frac{r_H \sin \frac{\pi}{5}}{r + r_H \cos \frac{\pi}{5}}  \cr
  & + \sqrt{5} \operatorname{artanh} \frac{\sqrt{5} r r_H}{2 r^2 + r r_H + 2 r_H^2} +\frac{5}{2} \ln \frac{r^4 + r^3 r_H + r^2 r_H^2 + r r_H^3 + r_H^4}{r^4} \bigg] , \quad \text{(p=4)} \cr
  F(r) =&\; \frac1{6r_H} \bigg[ 2\sqrt3 \arctan \frac{ \sqrt{3} rr_H }{r^2-r_H^2} \cr
  & + \ln\frac{(r+r_H^2)^4 (r^2 + r r_H + r_H^2)^2 (r^4 + r^2 r_H^2 + r_H^4)}{r^{12}} \bigg]. \quad \text{(p=5)}
\end{align}
Here we check the cases for $2\leq p \leq 5$. The background for the case of $p=1$ is the AdS$_3$ black hole which can not be compactified and has been discussed in \cite{Haack0806}. For $p \geq 6$, the differential equation \cref{eq: diff eq of 1st order alpha(ij)} can still be solved, but we omit the explicit solutions here---the cases of $2\leq p \leq 5$ are enough for us to get the general form of the results.

Then one can use the $(ri)$-component of Einstein equation $E_{ri}-T_{ri}=0$, to solve the vector perturbation as
\begin{align}\label{eq: 1st order w_i solution}
  w_i^{(1)}(r) = \frac{1}{r}\partial _0\beta_i,
\end{align}
and use the linear combination of $(0i)$- and $(ri)$-components $g^{r0} (E_{0i} - T_{0i}) + g^{rr} (E_{ri} - T_{ri}) = 0$ to get the first order vector constraint
\begin{align}\label{eq: 1st order vector constraint}
    \frac{1}{r_H} \partial _i r_H = - \partial _0 \beta_i.
\end{align}

For the scalar part, the first scalar constraint is
\begin{align}\label{eq: 1st order scalar constraint 1}
  \frac1{r_H} \partial_0 r_H = - \frac1{p} \partial \beta,
\end{align}
which is got through substituting the first order expanded metric \cref{eq: 1st order expanded metric} into $g^{r0} (E_{00} - T_{00}) + g^{rr} (E_{r0} - T_{r0}) = 0$. To solve the scalar perturbations, we use the combination of $(rr)$- and $(r0)$-components $g^{rr} (E_{rr} - T_{rr}) + g^{r0} (E_{r0} - T_{r0}) = 0$, the $(rr)$-component $E_{rr}-T_{rr} = 0$ and the EOM of the background scalar $A$ \cref{eq: EOM A} and get the differential equations for the first order scalar perturbations as
\begin{align}
  & (r^{p+1} k_{(1)})' + 2 (p+1) r^p j_{(1)} + \left[ (p-q) r^{p+1} - \frac{(p-q)(p-1)}{2p} r_H^{p+1} \right] h'_{(1)} + 2 r^{p-1} \partial \beta = 0, \label{eq: diff eq from 1st order scalar constraint2} \\
  & (p-q) r h''_{(1)} + 2(p-q) h'_{(1)} + 2p j'_{(1)} = 0, \label{eq: diff eq from 1st order EOM(rr)} \\
  & (r^{p+1} k_{(1)})' + r^{p+1} f j'_{(1)} + 2(p+1) r^p j_{(1)} + \frac{p-q}{2} r^{p+1} f h'_{(1)} + r^{p-1} \partial \beta = 0. \label{eq: diff eq from 1st order EOM dilaton}
\end{align}
The solutions are
\begin{align}\label{eq: 1st order hjk solutions}
  F_h &= \frac1{p-q} F, \qquad F_j = \frac{1}{p} \frac{ r^p-r_H^p }{ r^{p+1} - r_H^{p+1} } - \frac{1}{2p} F, \cr
  F_k &= - \frac2{pr} + \frac{1}{p} \left( 1+ \frac{(p-1) r_H^{p+1}}{2 r^{p+1}} \right) F.
\end{align}
Here we have $\psi = F_\psi \partial \beta$ $(\psi = \{h_{(1)},\ j_{(1)},\ k_{(1)}\})$.

The boundary stress-energy tensor can be got by calculating the following Brown-York tensor in large $r$ limit:
\begin{align}\label{eq: Brown-York tensor}
  T_{\mu\nu} = \frac{1}{\kappa_{p-q+2}^2} \lim_{r\to\infty} \left( \frac rL \right)^\frac{p(p-q-1)}{p-q} \left( K_{\mu\nu} - h_{\mu\nu}K - \frac{p}{L} \left( \frac rL \right)^{-\frac{q}{p-q}} h_{\mu\nu} \right).
\end{align}
Using the first order on-shell metric, we get the first order stress-energy tensor for the relativistic fluids that are dual to compactified AdS black holes
\begin{align}\label{eq: 1st order stress-energy tensor}
  T_{\mu\nu} =& \frac{1}{2 \kappa_{p-q+2}^2} \left[ \frac{r_H^{p+1}}{L^{p+2}} \left( p ~ u_\mu u_\nu + P_{\mu\nu} \right) - \left( \frac{r_H}{L} \right) ^p \left( 2\sigma_{\mu\nu} + \frac{2q}{p(p-q)} P_{\mu\nu} \partial u \right) \right],
\end{align}
from which we find
\begin{align}\label{eq: 1st order transport coefficients}
  \varepsilon &= \frac1{2 \kappa_{p-q+2}^2} \; p \; {r_H^{p+1} \over L^{p+2}}, \hspace{1.cm} \mathfrak p = \frac{1}{2 \kappa_{p-q+2}^2} {r_H^{p+1} \over L^{p+2}}, \cr
   \eta &= \frac{1}{2 \kappa_{p-q+2}^2} \left( \frac{r_H}{L} \right)^p, \hspace{1.cm} \zeta = \frac{1}{2 \kappa_{p-q+2}^2} \frac{2q}{p(p-q)} \left( \frac{r_H}{L} \right)^p.
\end{align}
From the result of $\zeta$, one can see that the numerical part of it is indeed the same as the $\gamma^2$ in \cref{eq: gamma^2 in CR form of comp. AdS BH}. Since the Hawking temperature of \cref{eq: (p-q+2) dimensional reduced metric of cAdS BH} is
\begin{align}
  T = \frac{(p+1) r_H}{4\pi L^2}.
\end{align}
One can get the entropy density and the ratios involved with it as
\begin{align}
  s = \frac{1}{2 \kappa_{p-q+2}^2} 4\pi \left( \frac{r_H}{L} \right)^p, \qquad \frac\eta s = \frac{1}{4\pi}, \qquad \frac\zeta s = \frac{1}{2\pi} \frac{q}{p(p-q)}.
\end{align}
One can also get the sound speed by
\begin{align}
  c_s^2 = \left(\frac{\partial \mathfrak p}{\partial \varepsilon}\right)_s = \frac{1}{p},
\end{align}
which is the same as the AdS black hole. We would also like to add some comments on the first-order results. First, one has
\begin{align}
  \frac\zeta\eta = \frac{2q}{p(p-q)} = 2 \left( \frac{1}{p-q} - c_s^2 \right),
\end{align}
which means that the relativistic fluids that are dual to compactified AdS black holes saturate the Buchel bound of bulk viscosity \cite{Buchel0708}. This is also true for the case of compactified Dp-brane \cite{Wu2012}. Second, the bulk viscosity offered here reproduces the result of CR$_5$ in \cite{Gubser0806} by setting $q=p-3$. Third, the bulk viscosity we get also coincides the result in \cite{Kanitscheider0901} by setting $2\sigma = p+1$ and $d = p-q+1$.

\section{The second-order results}

\subsection{Preliminaries and expansion of the on-shell metric}

We will be very brief in formulating the solving procedure for the second order since it is similar to the D(p-q)-brane \cite{Wu2012}. First, to solve the second-order, one needs the expressions of the second-order constraints derived through $\partial_\mu \partial^\rho T^{(0)}_{\rho\nu} = 0$:
\begin{align}
  \frac1{r_H} \partial _0^2 r_H + \frac1p \partial_0\partial \beta - \frac{p-1}{p} \partial_0\beta_i \partial_0\beta_i - \frac{1}{p^2} (\partial \beta)^2 = 0, \label{eq: 2nd order constraint (00) raw}  \\
  \frac{1}{r_H} \partial_i^2 r_H + \partial_0 \partial \beta - \partial_0\beta_i \partial_0\beta_i - \frac{1}{p} (\partial \beta)^2 + \partial_i \beta_j \partial _j\beta_i = 0, \label{eq: 2nd order constraint (ii) raw} \\
  \frac{1}{r_H} \partial _0\partial _i r_H + \partial _0^2\beta_i - \frac{2}{p} \partial _0\beta_i \partial \beta + \partial_0 \beta_j \partial _j\beta_i = 0, \label{eq: 2nd order constraint (0i) raw} \\
  \frac1{r_H} \partial _0 \partial _i r_H + \frac1p \partial_i \partial \beta - \frac1p \partial _0 \beta_i \partial \beta - \frac{p-1}{p} \partial_0 \beta_j \partial _i\beta_j = 0, \label{eq: 2nd order constraint (i0) raw} \\
  \partial _0 \Omega_{ij} - \frac{1}{p} \Omega_{ij} \partial \beta - \partial_k\beta_{[i} \partial_{j]}\beta_k = 0, \label{eq: 2nd order constraint [ij] raw} \\
  \frac{1}{r_H} \partial_i \partial _j r_H + \partial _0 \partial_{(i}\beta_{j)} - \partial_0\beta_i \partial_0\beta_j - \frac{1}{p} \partial_{(i}\beta_{j)} \partial\beta + \partial _k\beta _{(i} \partial _{j)} \beta _k = 0
  \label{eq: 2nd order constraint (ij) raw}
\end{align}
with $T^{(0)}_{\mu\nu}$ the hydrostatic part of the constitutive relation. All the viscous terms above are defined in \cref{tab: 2nd order spatial viscous terms}.
\begin{table}[h]
\centering
\begin{tabular}{|l|l|l|}
  \hline
  Scalars of $\mathrm{SO}(p-q)$                                 &                 Vectors of $\mathrm{SO}(p-q)$                      &\quad  Tensors of $\mathrm{SO}(p-q)$ \\ \hline\hline
 $\mathbf{s}_1=\frac1{r_H}\partial_0^2r_H$         & $\mathbf{v}_{1i} = \frac1{r_H} \partial_0\partial_i r_H$ & $\mathbf{t}_{1ij} = \frac1{r_H} \partial_i\partial_j r_H - \frac1{p-q} \delta_{ij} \mathbf{s}_3$ \\
  $\mathbf{s}_2 = \partial_0\partial_i\beta_i$           & $\mathbf{v}_{2i} = \partial_0^2\beta_i$       & $\mathbf{t}_{2ij} = \partial_0 \Omega_{ij}$ \\
  $\mathbf{s}_3 = \frac1{r_H}\partial_i^2r_H$         & $\mathbf{v}_{3i} = \partial_j^2\beta_i$        & $\mathbf{t}_{3ij} = \partial_0\sigma_{ij}$ \\
  $\mathfrak S_1 = \partial_0\beta_i\partial_0\beta_i$ & $\mathbf{v}_{4i}=\partial_j\Omega_{ij}$      & $\mathfrak T_{1ij} = \partial_0\beta_i\partial_0\beta_j - \frac1{p-q} \delta_{ij} \mathfrak S_1$ \\
  $\mathfrak S_3 = (\partial_i\beta_i)^2$                    & $\mathbf{v}_{5i} = \partial_j\sigma_{ij}$      & $\mathfrak T_{2ij} = \sigma_{[i}^{~~k} \Omega_{j]k}$ \\
  $\mathfrak S_4 = \Omega_{ij} \Omega_{ij}$           & $\mathfrak V_{1i} = \partial_0\beta_i\partial\beta$     & $\mathfrak T_{3ij} = \Omega_{ij} \partial \beta$ \\
  $\mathfrak S_5=\sigma_{ij}\sigma_{ij}$                 &  $\mathfrak V_{2i} = \partial_0\beta_j \Omega_{ij}$  & $\mathfrak T_{4ij}=\sigma_{ij}\partial\beta$ \\
                                                                                        & $\mathfrak V_{3i} = \partial_0\beta_j \sigma_{ij}$  & $\mathfrak T_{5ij} = \Omega_i^{~k}\Omega_{jk} - \frac1{p-q} \delta_{ij} \mathfrak S_4$ \\
                                                                                        &                                                             & $\mathfrak T_{6ij} = \sigma_i^{~k}\sigma_{jk} - \frac1{p-q} \delta_{ij} \mathfrak S_5$ \\
                                                                                        &                                                             & $\mathfrak T_{7ij} = \sigma_{(i}^{~~k} \Omega_{j)k}$ \\
\hline
\end{tabular}
\caption{\label{tab: 2nd order spatial viscous terms} All the $\mathrm{SO}(p-q)$ invariant second-order viscous terms for the relativistic fluid which is dual to compactified AdS black hole. Here $p\geq 2$ and $1\leq q\leq p-1$.}
\end{table}
We may also re-express all the constraint relations of \cref{eq: 2nd order constraint (00) raw,eq: 2nd order constraint (0i) raw,eq: 2nd order constraint (0i) raw,eq: 2nd order constraint (i0) raw,eq: 2nd order constraint (ii) raw,eq: 2nd order constraint (ij) raw,eq: 2nd order constraint [ij] raw} as
\begin{align}
  & \mathbf s_1 + \frac1p \mathbf s_2 - \frac{p-1}{p} \mathfrak S_1 - \frac{1}{p^2} \mathfrak S_3 = 0, \label{eq: 2nd order constraint (00)} \\
  & \mathbf s_2 + \mathbf s_3 - \mathfrak S_1 + \frac{q}{p(p-q)} \mathfrak S_3 - \mathfrak
  S_4 + \mathfrak S_5 = 0, \label{eq: 2nd order constraint (ii)} \\
  & \mathbf v_1 + \mathbf v_2 - \frac{p-2q}{p(p-q)} \mathfrak V_1 - \mathfrak V_2 + \mathfrak V_3 = 0, \label{eq: 2nd order constraint (0i)} \\
  & \mathbf v_1 + \frac{p-q}{p(p-q-1)}(\mathbf v_4 + \mathbf v_5) - \frac{2p-q-1}{p(p-q)} \mathfrak V_1 - \frac{p-1}{p} (\mathfrak V_2 + \mathfrak V_3) = 0, \label{eq: 2nd order constraint (i0)} \\
  & \mathbf t_2 - 2 \mathfrak T_2 + \frac{p+q}{p(p-q)} \mathfrak T_3 = 0, \label{eq: 2nd order constraint [ij]} \\
  & \mathbf t_1 + \mathbf t_3 - \mathfrak T_1 + \frac{p+q}{p(p-q)} \mathfrak T_4 - \mathfrak T_5 + \mathfrak T_6 = 0. \label{eq: 2nd order constraint (ij)}
\end{align}

From the conservation equation $\partial^\mu T^{(0+1)}_{\mu\nu} = 0$, one will get the Navier-Stokes equations as
\begin{align}
  \partial_0 r_H^{(1)} =&\; \frac{2q}{p^2(p+1)(p-q)} \mathfrak S_3 + \frac{2}{p(p+1)} \mathfrak S_5, \label{eq: Navier-Stokes 0}\\
  \partial_i r_H^{(1)} =&\; \frac{2q~ \mathbf v_4 + 2(p-q)(p-1) \mathbf v_5}{p(p+1)(p-q-1)}+ \frac{(p-1)(p-2q+2) - pq}{p(p-q)(p+1)} \mathfrak V_1 \cr
 & - \frac{p+2}{p(p+1)} \mathfrak V_2 - \frac{2p^2 - 3p + 2}{p(p+1)} \mathfrak V_3.
 \label{eq: Navier-Stokes i}
\end{align}

Then we expand the global on-shell metric \cref{eq: global on-shell metric} to second-order and get
\begin{align}\label{eq: 2nd order expanded metric}
  ds^2 =& -r^\frac{2p}{p-q} \bigg[ f - (1-f) \delta\beta_i \delta\beta_i - \frac{(p+1)r_H^{p}}{r^{p+1}}(\delta r_H +\frac{1}{2} \delta^2r_H+\delta r_H^{(1)}) - \frac{p(p+1) r_H^{p-1}}{2r^{p+1}}(\delta r_H)^2 \cr
  & + (F_k+\delta F_k)\partial \beta + F_k(\delta\partial \beta + \delta\beta_i \partial_0\beta_i) + \frac{2}{r} \delta\beta_i \partial_0\beta_i + k^{(2)}(r) \bigg] dv^2 \cr
  &  + 2r^\frac{2p}{p-q} \bigg[ (f-1)(\delta\beta_i + \frac12 \delta^2\beta_i) + \frac{1}{r}(\partial_0\beta_i + \delta\partial_0\beta_i + \delta\beta_j\partial_j\beta_i) - \frac{(p+1)r_H^{p}}{r^{p+1}}\delta r_H \delta\beta_i \cr
  & + F_k \partial \beta \delta\beta_i - F \delta\beta_j \partial_{(i}\beta_{j)} + w_i^{(2)}(r) \bigg] dvdx^i + 2r^\frac{2q}{p-q} \bigg[ 1 - (F_j+\delta F_j) \partial \beta \cr
  & - F_j(\delta\partial \beta + \delta\beta_i\partial_0\beta_i) + \frac12 \delta\beta_i\delta\beta_i - j^{(2)}(r) \bigg] dvdr + r^\frac{2p}{p-q} \bigg[ \delta_{ij} + (1-f)\delta\beta_i\delta\beta_j \cr
  & - \frac{2}{r} \delta\beta_{(i}\partial_{|0|}\beta_{j)} + (F + \delta F) \partial_{(i}\beta_{j)} + F \left( \delta\partial_{(i} \beta_{j)} + \delta\beta_{(i} \partial_{|0|}\beta_{j)} \right) + \alpha_{ij}^{(2)}(r)\cr
  & + h^{(2)}(r) \delta_{ij} \bigg] dx^idx^j  - 2 r^\frac{2q}{p-q} \bigg( \delta\beta_i + \frac12 \delta^2\beta_i - F_j \partial\beta \delta\beta_i \bigg) dx^i dr.
\end{align}
Here we define
\begin{align}
  \delta \mathcal F(r_H(x),r) = - \frac{\mathcal{F}(r) + r \mathcal{F}'(r)}{r_H} \delta r_H.
\end{align}
Then put the second-order expanded on-shell metric into the Einstein equation and the dilaton's EOM one can get the differential equations for all the second-order perturbations. They can be solved by the same token as in \cite{Wu1604,Wu1807,Wu2012}. The details will be omitted in this paper.

\subsection{The dynamical second-order transport coefficients}

From the second-order on-shell metric, we get the stress-energy tensor of the relativistic fluids that are dual to $(p-q+2)$-dimensional compactified AdS black hole, or the background of CR$_{p-q+2}$:
\begin{align}\label{eq: CAdS(p-q+2) 2nd order stress-energy tensor}
  T_{\mu\nu} =&\; \frac{1}{2 \kappa_{p-q+2}^2} \Bigg\{ {r_H^{p+1} \over L^{p+2}} \left( p~ u_\mu u_\nu + P_{\mu\nu} \right) - \left( \frac{r_H}{L} \right) ^p \bigg( 2\sigma_{\mu\nu} + \frac{2q}{p(p-q)} P_{\mu\nu} \partial u \bigg) \cr
  & + \frac{r_H^{p-1}}{L^{p-2}} \Bigg[ \bigg( \frac12 + \frac{1}{p+1} H_\frac{2}{p+1} \bigg)\cdot 2\bigg( \sideset{_\langle}{}{\mathop D}\sigma_{\mu\nu\rangle} + \frac1{p-q} \sigma_{\mu\nu} \partial u \bigg) \cr
  & + \bigg( \frac{3q}{2p} - \frac{q}{p(p+1)} H_\frac{2}{p+1} \bigg) \frac{2\sigma_{\mu\nu} \partial u}{p-q} + \frac{1}{2} \cdot 4\sigma_{\langle\mu}^{~~\rho}\sigma_{\nu\rangle\rho} + \bigg( - 1 + \frac{2}{p+1} H_\frac{2}{p+1} \bigg) \cdot 2 \sigma_{\langle\mu}^{~~\rho} \Omega_{\nu\rangle\rho} \Bigg] \cr
  & + P_{\mu\nu} \frac{r_H^{p-1}}{L^{p-2}} \Bigg[ \bigg( \frac{q}{p(p-q)} + \frac{2q}{p(p+1)(p-q)} H_\frac{2}{p+1} \bigg) D(\partial u) \cr
  & + \Bigg( \frac{q(3q-p)}{p^2(p-q)^2} + \frac{2q}{p^2(p+1)(p-q)} H_\frac{2}{p+1} \Bigg) (\partial u)^2 + \frac{q}{2p(p-q)} \cdot 4 \sigma_{\alpha\beta}^2 \Bigg] \Bigg\}.
\end{align}
Then we can read all the second-order transport coefficients as
\begin{align}\label{eq: compactified AdS-BH 2nd order coefficients}
  \eta\tau_\pi &= \frac1{2\kappa_{p-q+2}^2} \left( \frac12 + \frac{1}{p+1} H_\frac{2}{p+1} \right) \frac{r_H^{p-1}}{L^{p-2}}, \cr
  \eta\tau_\pi^* &= \frac1{2\kappa_{p-q+2}^2} \left[ \frac{3q}{2p} - \frac{q}{p(p+1)} H_\frac{2}{p+1} \right] \frac{r_H^{p-1}}{L^{p-2}}, \cr
  \lambda_1 &= \frac1{2\kappa_{p-q+2}^2} \frac12 \frac{r_H^{p-1}}{L^{p-2}}, \qquad \lambda_2 = \frac1{2\kappa_{p-q+2}^2} \left( - 1 + \frac{2}{p+1} H_\frac{2}{p+1} \right) \frac{r_H^{p-1}}{L^{p-2}}, \cr
  \zeta\tau_\Pi &= \frac1{2\kappa_{p-q+2}^2} \Bigg[ \frac{q}{p(p-q)} + \frac{2q}{p(p+1)(p-q)} H_\frac{2}{p+1} \Bigg] \frac{r_H^{p-1}}{L^{p-2}}, \cr
  \xi_1 &= \frac1{2\kappa_{p-q+2}^2} \frac{q}{2p(p-q)} \frac{r_H^{p-1}}{L^{p-2}}, \cr
  \xi_2 &= \frac1{2\kappa_{p-q+2}^2} \Bigg[ \frac{q(3q-p)}{p^2(p-q)^2} + \frac{2q}{p^2(p+1)(p-q)} H_\frac{2}{p+1} \Bigg] \frac{r_H^{p-1}}{L^{p-2}}.
\end{align}
Here $H_\frac{2}{p+1}$ is the harmonic number and it is defined as
\begin{align}
  H_\frac ab = \frac ba + 2 \sum_{n=1}^{\left[ \frac{b-1}{2} \right]} \cos\left( \frac{2\pi n a}{b} \right) \ln\sin\left( \frac{n\pi}{b} \right) - \frac{\pi}{2} \cot\left( \frac{\pi a}{b} \right) - \ln(2b),
\end{align}
with $a$ and $b$ are both integers. $\left[ \frac{b-1}{2} \right]$ means taking the integer part of $\frac{b-1}{2}$. For the lowest values of $p$, the above gives
\begin{align}
  H_{\frac23} &= \frac32 + \frac{\pi}{2\sqrt3} - \frac32 \ln3, \qquad (p=2) \\
  H_{\frac12} &= 2 - 2 \ln2, \qquad (p=3) \\
  H_{\frac25} &= \frac52 - \frac\pi2 \sqrt{1 - \frac2{\sqrt5}} + \frac{\sqrt5}{2} \operatorname{arcoth} \sqrt5 - \frac54 \ln5, \quad (p=4) \\
  H_{\frac13} &= 3 - \frac{\pi}{2\sqrt3} - \frac32 \ln3. \qquad (p=5)
\end{align}
The second-order transport coefficients as offered in \cref{eq: compactified AdS-BH 2nd order coefficients} can reproduce the dynamical ones in \cite{Kleinert1610} by setting $q=p-3$.\footnote{For direct comparison with the results in \cite{Kleinert1610}, one may need an identity about the Harmonic number as $H_{\alpha-1} = H_\alpha - \frac{1}{\alpha}$}

From \cref{eq: compactified AdS-BH 2nd order coefficients}, we can find that of the 7 dynamical second-order transport coefficients, only $\eta \tau_\pi^*,\ \zeta \tau_\Pi,\ \xi_1$, and $\xi_2$ depend on $q$, which is the same as D(p-q)-brane \cite{Wu2012}. As we have explained in \cite{Wu2012}, these $q$-depending coefficients are also non-conformal. They all relate to the viscous term $\partial u$, which is the trace part of shear viscous tensor $\sigma_{\mu\nu}$, and compactifying spatial dimensions will diminish the components that are summed in taking  trace. That's why only $\eta \tau_\pi^*,\ \zeta \tau_\Pi,\ \xi_{1,2}$ depend on the $q$, i.e. the number of directions that are compactified.

The identities among the dynamical coefficients that hold true for D(p-q)-brane \cite{Wu1604,Wu1807,Wu2012} are still valid for compactified AdS black holes. We list these identities here for later convenience:
\begin{align}
  & \text{Haack-Yarom relation:} \qquad 4\lambda_1 + \lambda_2 = 2 \eta \tau_\pi, \\
  & \text{Romatschke relation 1:} \qquad \tau_\pi = \tau_\Pi, \\
  & \text{Romatschke relation 2:} \qquad \xi_1 = \frac{1}{p-q} [1 - (p-q) c_s^2] \lambda_1, \\
  & \text{Kleinert-Probst relations:} \qquad \eta\tau^*_\pi = \left(1 - (p-q) c_s^2 \right) (4\lambda_1 - \eta\tau_\pi), \qquad  \cr
  & \xi_2 = \frac2{(p-q)^2} \left(1 - (p-q) c_s^2 \right) \left[ (1 - 2(p-q) c_s^2) 2\lambda_1 + (p-q) c_s^2 \eta\tau_\pi \right]. \label{eq: Kleinert-Probst relations}
\end{align}
As we have pointed out in \cite{Wu2012} that these 5 identities reduce the independent dynamical coefficients from 7 to 2, which can be chosen as $\tau_\pi$ and $\lambda_1$.

The dispersion relations can be calculated by the same method as in \cite{Wu1604,Wu1807,Wu2012}, the result is
\begin{align}
  \omega_T =& -i \frac{1}{p+1} \frac{L^2}{r_H} k^2 - i \, \frac{p+1+2H_{\frac{2}{p+1}}}{2(p+1)^3} \frac{L^6}{r_H^3} k^4, \\
  \omega_L =& \pm \frac{1}{\sqrt{p}} k - i \frac{p-1}{p(p+1)} \frac{L^2}{r_H} k^2 \pm \frac{(p-1) \left(1 + H_\frac{2}{p+1} \right)}{p^{3/2} (p+1)^2} \frac{L^4}{r_H^2} k^3 \cr
  &- i \frac{(p-1)^2 \left( p+1+2H_\frac{2}{p+1} \right)}{p^2 (p+1)^3} \frac{L^6}{r_H^3} k^4.
\end{align}
The dispersion relations do not depend on $q$, which is the same as the compactified Dp-brane \cite{Wu2012}. The compactification does not affect the dispersion relations, as it does for the sound speed.

\subsection{The second-order thermodynamical transport coefficients}

In this subsection, we will offer the results for the second-order thermodynamical transport coefficients. But they are not derived by direct calculations as what we have done for the 7 dynamical ones. Thus the results need to be checked via other methods like the Minkowskian AdS/CFT correspondence \cite{Baier0712,Barnes1004,Arnold1105}.

The second-order transport coefficients of the 5-dimensional compactified AdS black hole (or CR$_5$) have been given in \cite{Bigazzi1006} and \cite{Kleinert1610}. \cite{Bigazzi1006} gives the results to the first order of $\delta_\text{nonconf.}$ expansion, whose definition in 5 dimensions is
\begin{align}
  \delta_\text{nonconf.} = 1-3c_s^2 = 1 - \frac{3}{p}.
\end{align}
\cite{Kleinert1610} offers the exact results which are also formulated in terms of $\delta_\text{nonconf.}$. Since in general dimensions, we have
\begin{align}
  \delta_\text{nonconf.} = 1 - (p-q) c_s^2 = 1 - \frac{q}{p},
\end{align}
which means if we substitute factors involved with $\delta_\text{nonconf.}$ for the form in general dimensions, we can generalize the thermodynamical coefficients to other dimensions. According to \cite{Bigazzi1006,Kleinert1610}, the coefficients $\lambda_{3,4}$ and $\xi_{3,4}$ of CR$_5$ are all 0, we can get 4 non-trivial thermodynamical coefficients as
\begin{align}
  \kappa &= \frac1{2\kappa_{p-q+2}^2} \frac{2}{p-1} \frac{r_H^{p-1}}{L^{p-2}}, \hspace{6.15em} \kappa^* = - \frac1{2\kappa_{p-q+2}^2} \frac{q}{p-1} \frac{r_H^{p-1}}{L^{p-2}}, \cr
  \xi_5 &= \frac1{2\kappa_{p-q+2}^2} \frac{2q}{p(p-1)(p-q)} \frac{r_H^{p-1}}{L^{p-2}}, \qquad \xi_6 = \frac1{2\kappa_{p-q+2}^2} \frac{2q}{(p-1)(p-q)} \frac{r_H^{p-1}}{L^{p-2}},
  \label{eq: compactified AdS-BH thermal coefficients}
\end{align}
which satisfy the following 3 identities
\begin{align}\label{eq: relations of kappa* xi5 xi6 with kappa}
  \kappa^* = - \frac{1-(p-q) c_s^2}{2 c_s^2} \kappa, \qquad \xi_5 = \frac{1-(p-q) c_s^2}{p-q} \kappa, \qquad \xi_6 = \frac{1-(p-q) c_s^2}{(p-q) c_s^2} \kappa.
\end{align}
These identities are firstly proposed in \cite{Romatschke0906} for 4-dimensional relativistic fluids. Then \cite{Wu1807} makes a guess on their forms in general dimensions. By replacing $p-q$ with $d-1$, where $d$ is the number of dimensions of the dual fluids, one can see that the above identities are in accord with the forms proposed in \cite{Wu1807}. For a comparison between CR$_5$ and CR$_{p-q+2}$, we list all the non-trivial transport coefficients up to the second-order of these two cases in \cref{tab: comparison between CR5 and CR(p-q+2)}. The data of CR$_5$ comes from \cite{Kleinert1610}, we just reformulate their results in our convention.
\begin{table}[h]
\centering
\begin{tabular}{|c|c|c|}
    \hline
    Coefficient  &       CR$_5$                                                    &    CR$_{p-q+2}$ \\ [.3em] \hline \hline
    $\eta$       & $1$                                                                   &  $1$ \\ \hline
    $\zeta$      & $\frac{2(p-3)}{3p}$                                                   & $\frac{2q}{p(p-q)}$ \\ [.3em] \hline
    $\eta \tau_\pi$   & $\frac{1}{2}+\frac{1}{p+1}H_{\frac{2}{p+1}}$                          & $\frac{1}{2}+\frac{1}{p+1}H_{\frac{2}{p+1}}$ \\ [.3em] \hline
    $\kappa$     & $\frac{2}{p-1}$                                                       & $\frac{2}{p-1}$ \\ [.3em] \hline
    $\lambda_1$  & $\frac{1}{2}$                                                         & $\frac{1}{2}$ \\ [.3em] \hline
    $\lambda_2$  & $-1+\frac{2}{p+1}H_{\frac{2}{p+1}}$                                   & $-1+\frac{2}{p+1}H_{\frac{2}{p+1}}$ \\ [.3em]\hline
    $\lambda_3$  & $0$                                                                   & $0$ \\ \hline
    $\eta \tau_\pi^*$ & $\frac{3(p-3)}{2p}-\frac{p-3}{p(p+1)}H_{\frac{2}{p+1}}$               & $\frac{3q}{2p}-\frac{q}{p(p+1)}H_{\frac{2}{p+1}}$ \\ [.3em]\hline
    $\kappa^*$   & $- \frac{p-3}{p-1}$                                                     & $-\frac{q}{p-1}$ \\ [.3em] \hline
    $\lambda_4$  & $0$                                                                   & $0$ \\ \hline
    $\zeta \tau_\Pi$   & $\frac{p-3}{3p} + \frac{2(p-3)}{3p (p+1)} H_{\frac{2}{p+1}}$                          & $\frac{q}{p(p-q)} + \frac{2q}{p (p+1) (p-q)} H_{\frac{2}{p+1}}$ \\ [.3em] \hline
    $\xi_1$      & $\frac{p-3}{6p}$                                                      & $\frac{q}{2p(p-q)}$ \\ [.3em] \hline
    $\xi_2$      & $\frac{(p-3)(2p-9)}{9p^2}+\frac{2(p-3)}{3p^2(p+1)}H_{\frac{2}{p+1}}$ & $\frac{q(3q-p)}{p^2(p-q)^2}+\frac{2q}{p^2(p+1)(p-q)}H_{\frac{2}{p+1}}$ \\ [.3em] \hline
    $\xi_3$      & $0$                                                                   & $0$ \\ \hline
    $\xi_4$      & $0$                                                                   & $0$ \\ \hline
    $\xi_5$      & $\frac{2(p-3)}{3p(p-1)}$                                              & $\frac{2q}{p(p-1)(p-q)}$ \\ [.3em] \hline
    $\xi_6$      & $\frac{2(p-3)}{3(p-1)}$                                               & $\frac{2q}{(p-1)(p-q)}$  \\ [.3em] \hline
\end{tabular}
\caption{\label{tab: comparison between CR5 and CR(p-q+2)} The comparison between the transport coefficients of CR$_5$ and CR$_{p-q+2}$, here the Chamblin-Reall backgrounds refer to the compactified AdS black hole.}
\end{table}

The 8 second-order thermodynamical coefficients satisfy 5 differential constraint relations firstly found in  \cite{Bhattacharyya1201} by Bhattacharyya, leaving 3 of them independent, which have been chosen to be $\kappa$ and $\lambda_{3,4}$ in \cite{Wu2012}. These 5 constraint relations, which we will call the Bhattacharyya relations, are formulated in a different but more convenient form in \cite{Moore1210}. Yet, this form does not reflect the independence of $\kappa$ and $\lambda_{3,4}$. We rewrite the Bhattacharyya relations for 4-dimensional relativistic fluids in a form that emphasizes the independent coefficients here:
\begin{align}
  \kappa^* &= - \left( \frac12 T \partial_T - 1 \right) \kappa, \\
  \xi_5 &= \frac16 \left[ 3 c_s^2 T \partial_T - (1+3c_s^2) \right] \kappa, \\
  \xi_6 &= \left[ c_s^2 T^2 \partial_T^2 - \frac13 (2-3c_s^2) T \partial_T + \frac13 (1-3c_s^2) \right] \kappa + \frac{\lambda_4}{c_s^2}, \\
  \xi_3 &= \frac{1}{12} \left[ 3c_s^2 T \partial_T + (1-9c_s^2) \right] \lambda_3 - \frac34 \left[ c_s^2 T^2 \partial_T^2 - (1-2c_s^2) T \partial_T \right] \kappa - \frac{\lambda_4}{c_s^2}, \\
  \xi_4 &= - \frac16 \left[ 3c_s^2 T \partial_T + (1+3c_s^2) \right] \lambda_4 - \frac12 c_s^4 \left[ c_s^2 T^3 \partial_T^3 - (1-3c_s^2) T^2 \partial_T^2 \right] \kappa.
\end{align}
Since in the compactified AdS black hole, the temperature dependence of the second-order transport coefficients is $\sim r_H^{p-1} \sim T^{p-1}$. Then one has the following equivalences:
\begin{align}
  T \partial_T \Leftrightarrow p-1, \qquad T^2 \partial_T^2 \Leftrightarrow (p-1)(p-2), \qquad T^3 \partial_T^3 \Leftrightarrow (p-1)(p-2)(p-3).
\end{align}
In the Chamblin-Reall backgrounds with only one background scalar, the temperature dependence of the transport coefficients is just power law, thus the differential relations in Bhattacharyya constraints can be reduced to algebraic identities like the Romatschke or Kleinert-Probst relations. This is indeed true as it can be checked by putting the results of \cref{eq: compactified AdS-BH thermal coefficients} into the 4-dimensional Bhattacharyya relations in the special case of $q=p-3$. But how about in general dimensions?

We make a bold guess on the Bhattacharyya constraints about their forms in general dimensions, which are
\begin{align}
  \kappa^* =& - \frac12 \left( T \partial_T - (p-q-1) \right) \kappa, \\
  \xi_5 =&\; \frac1{2(p-q)} \left[ (p-q) c_s^2 T \partial_T - \left( (p-q-2) + (p-q) c_s^2 \right) \right] \kappa, \\
  \xi_6 =& \left[ c_s^2 T^2 \partial_T^2 - \frac{(p-q-1) - (p-q) c_s^2}{p-q} T \partial_T + \frac{1 - (p-q) c_s^2}{p-q} \right] \kappa + \frac{\lambda_4}{c_s^2}, \\
  \xi_3 =&\; \frac{(p-q) c_s^2 T \partial_T + \left( 1 - (p-q)^2 c_s^2 \right)}{(p-q) (p-q-1)} \lambda_3 - \frac{p-q}{p-q+1} \left[ c_s^2 T^2 \partial_T^2 - (1-2c_s^2) T \partial_T \right] \kappa - \frac{\lambda_4}{c_s^2},
  \label{eq: Bhattacharyya constraints xi3} \\
  \xi_4 =& - \frac{2 \left[ (p-q) c_s^2 T \partial_T + (1 + (p-q) c_s^2) \right]}{(p-q) (p-q-1)} \lambda_4 - \frac12 c_s^4 \left[ c_s^2 T^3 \partial_T^3 - (1-3c_s^2) T^2 \partial_T^2 \right] \kappa.
  \label{eq: Bhattacharyya constraints xi4}
\end{align}
We have checked the above for the case of compactified AdS black hole with $\lambda_{3,4}=\xi_{3,4}=0$. The relations of $\kappa^*,\ \xi_5$, and $\xi_6$ can indeed be reduced to algebraic forms of \cref{eq: relations of kappa* xi5 xi6 with kappa} by using \cref{eq: compactified AdS-BH thermal coefficients}. The factor $\left[ c_s^2 T^2 \partial_T^2 - (1-2c_s^2) T \partial_T \right]$ in \cref{eq: Bhattacharyya constraints xi3} and $\left[ c_s^2 T^3 \partial_T^3 - (1-3c_s^2) T^2 \partial_T^2 \right]$ in \cref{eq: Bhattacharyya constraints xi4} are automatically zero for 5-dimensional compactified AdS black hole, i.e. when $q = p-3$. We can not find their forms in $p-q+2$ dimensions that can still automatically vanish, so we just leave them as they are. One can check that when $\lambda_4=0$ and ignoring the term $\left[ c_s^2 T^2 \partial_T^2 - (1-2c_s^2) T \partial_T \right]$, \cref{eq: Bhattacharyya constraints xi3} will reduce to the following algebraic form
\begin{align}\label{eq: constraint xi3 and lambda3}
  \xi_3 = \frac{1 - (p-q) c_s^2}{p-q} \lambda_3,
\end{align}
which is the same as in \cite{Wu1807} given that $p-q = d-1$. We can also find that if ignoring the term of $\kappa$, \cref{eq: Bhattacharyya constraints xi4} will lead to a constraint relation between $\xi_4$ and $\lambda_4$ as
\begin{align}\label{eq: constraint xi4 and lambda4}
  \xi_4 = - \frac{2}{p-q} \lambda_4.
\end{align}
We hope that future studies will recover whether this is correct.

By now, we have 5 constraints among the thermodynamical second-order transport coefficients: \cref{eq: relations of kappa* xi5 xi6 with kappa,eq: constraint xi3 and lambda3,eq: constraint xi4 and lambda4}.
\cref{eq: relations of kappa* xi5 xi6 with kappa,eq: constraint xi3 and lambda3} are just the Romatschke relations \cite{Romatschke0906} in general dimensions, and \cref{eq: constraint xi4 and lambda4} is newly proposed in this work. These 5 algebraic relations are just the 5 Bhattacharyya constraints \cite{Bhattacharyya1201} when the transport coefficients have a simple power-law with respect to temperature. Such case is true for the Chamblin-Reall backgrounds with only one background scalar like the D(p-q)-brane and the compactified AdS black hole. When the dependence on temperature of the coefficients becomes complicated, these 5 algebraic relations \cref{eq: relations of kappa* xi5 xi6 with kappa,eq: constraint xi3 and lambda3,eq: constraint xi4 and lambda4} should be replaced with their more involved differential forms: the Bhattacharyya constraint relations.

In \cite{Kleinert1610}, the authors proposed another relation similar as the Haack-Yarom's, which reads $2(\kappa - \kappa^*) + \lambda_2 = 2 \eta \tau_\pi$. But using the results in \cref{eq: compactified AdS-BH 2nd order coefficients} and \cref{eq: compactified AdS-BH thermal coefficients} one can get
\begin{align}\label{eq: Kleiner-Probst new constraint}
  2(\kappa - \kappa^*) + \lambda_2 - 2 \eta \tau_\pi = - \frac{2 (p-q-3)}{p-1},
\end{align}
from which we can see that it is only held in 5-dimensions, i.e. $q=p-3$, as a special case. When $q$ goes general, the right-hand side of the above is no longer zero. So if $\kappa^*$ that we get is correct, this relation may not be valid in general dimensions.

We may explain the non-validity of \cref{eq: Kleiner-Probst new constraint} from another prospect. We have known from \cite{Wu2012} that only 5 of the 15 second-order transport coefficients are actually independent: the dynamical transport coefficients $\tau_\pi,\, \lambda_1$, and the thermodynamical ones $\kappa,\, \lambda_{3,4}$. If \cref{eq: Kleiner-Probst new constraint} holds, another one of these 5 would be taken from the list. As these remaining 5 have completely different physical meanings:
\begin{align}
  & \tau_\pi:\ \text{relaxation time} \qquad \lambda_1:\ \text{second-order shear viscosity} \qquad \kappa:\ \text{curvature} \cr
  & \lambda_3:\ \text{vorticity} \hspace{4.9em} \lambda_4:\ \text{gradient of entropy density}
\end{align}
One may infer that these 5 should all exist in the list of independent coefficients. So the conclusion may be \cref{eq: Kleiner-Probst new constraint} not hold in general dimensions. For the sake of prudence, one needs to use the results of the thermal coefficients by direct calculation to check.

Another observation on \cref{eq: Kleiner-Probst new constraint} is that it is a constraint among the transport coefficients that from both the dynamical and thermal sectors: $\tau_\pi,\, \lambda_2$ are from the dynamical sector while $\kappa, \, \kappa^*$ are thermal coefficients. The other constraints are among the coefficients within only one sector, either dynamical or thermal. This fact puts a question to us whether there should be a constraint that is among the coefficients belonging to different sectors? We hope future works may find a satisfying answer.

\section{A revisit to Kanitscheider-Skenderis proposal}

The Kanitscheider-Skenderis (KS) proposal \cite{Kanitscheider0901} allows one to extract the second-order transport coefficients for a non-conformal relativistic fluid from a conformal one in higher dimensions. This is realized by making substitutions in the conformal stress-energy tensor of higher dimensions. Under the convention of this work, the substitution rules proposed in \cite{Kanitscheider0901} can be rewritten as
\begin{align}
  \sigma_{\mu\nu} &\to \sigma_{\mu\nu} + \chi P_{\mu\nu} \partial u, \label{eq: substitution of sigma in KS proposal} \\
  A_{\langle\mu\nu\rangle} &\to  A_{\langle\mu\nu\rangle} + \chi P_{\mu\nu} P^{\rho\lambda} A_{\rho\lambda}, \label{eq: substitution of A in KS proposal} \\
  \Omega_{\mu\nu} &\to \Omega_{\mu\nu}, \quad R_{\rho\langle\mu\nu\rangle\sigma} \to R_{\rho\langle\mu\nu\rangle\sigma}, \quad R_{\langle\mu\nu\rangle} \to R_{\langle\mu\nu\rangle}, \quad R \to R.
  \label{eq: substitution of thermal sector in KS proposal}
\end{align}
The viscous terms of the thermodynamical sector do not change, only dynamical sector changes. $\chi$ is a parameter which contains both the number of dimensions of the conformal fluid $\tilde d$ and that of the non-conformal one $d$, and it is defined as
\begin{align}
  \chi = \frac{\tilde d - d}{(\tilde d - 1) (d-1)}.
\end{align}
For the case of compactified AdS black hole, $\tilde d = p+1$ and $d = p-q+1$. Such that
\begin{align}\label{eq: q-dependence of chi}
  \chi = \frac{q}{p (p-q)} = \frac{1-(p-q)c_s^2}{p-q}.
\end{align}
Since both the lower dimensional, non-conformal fluid and the conformal one in higher dimensions have dual gravitational backgrounds. These two backgrounds should be connected by dimensional reduction.
From the above introduction on the KS proposal, one can see that its spirit is precisely consistent with what we do for the compactified AdS black hole in this work.

\subsection{The dynamical sector}

Based on the rules from \cref{eq: substitution of sigma in KS proposal,eq: substitution of A in KS proposal,eq: substitution of thermal sector in KS proposal}, one can infer the substitution rules for the following viscous tensors in dynamical sector:
\begin{align}
  \sideset{_\langle}{}{\mathop D} \sigma_{\mu\nu\rangle}, \qquad \sigma_{\mu\nu} \partial u, \qquad \sigma_{\langle\mu}^{~~\rho}\sigma_{\nu\rangle\rho}, \qquad \sigma_{\langle\mu}^{~~\rho}\Omega_{\nu\rangle\rho}.
\end{align}
The KS proposal does not tell us how exactly the above viscous tensors will change. We need to figure out the transformation rules one by one with great deliberation. The first is $\sideset{_\langle}{}{\mathop D} \sigma_{\mu\nu\rangle}$. If we use the rule of \cref{eq: substitution of A in KS proposal} for it, we will find it does not change, since $P^{\rho\sigma} D \sigma_{\rho\sigma} = 0$. By using \cref{eq: substitution of sigma in KS proposal}, after some calculations, one can get
\begin{align}
  \sideset{_\langle}{}{\mathop D}\sigma_{\mu\nu\rangle} \to \sideset{_\langle}{}{\mathop D}\sigma_{\mu\nu\rangle} +\chi P_{\mu\nu} D(\partial u).
\end{align}
The transformation rule of $\sigma_{\mu\nu} \partial u$ can also be easily got with \cref{eq: substitution of sigma in KS proposal}:
\begin{align}
  \sigma_{\mu\nu}\partial u \to \sigma_{\mu\nu}\partial u +\chi P_{\mu\nu}(\partial u)^2.
\end{align}
Under either \cref{eq: substitution of sigma in KS proposal} or \cref{eq: substitution of A in KS proposal}, we have
\begin{align}
  \sigma_{\langle\mu}^{~~\rho} \Omega_{\nu\rangle\rho} \to \sigma_{\langle\mu}^{~~\rho} \Omega_{\nu\rangle\rho}.
\end{align}
One can check that the above three substitution rules are the same as \cite{Wu1807}. The last and most puzzled one is $\sigma_{\langle\mu}^{~~\rho}\sigma_{\nu\rangle\rho}$. If we use the rule of \cref{eq: substitution of A in KS proposal} we can only get $\sigma_{\langle \mu}^{~~\rho}\sigma_{\nu \rangle \rho} \to \sigma_{\langle \mu}^{~~\rho}\sigma_{\nu \rangle \rho} + \chi P_{\mu\nu} \sigma_{\alpha \beta}^2$, the same as in \cite{Wu1807}. But it is wrong! The correct rule for $\sigma_{\langle \mu}^{~~\rho}\sigma_{\nu \rangle \rho}$ is \cref{eq: substitution of sigma in KS proposal}. After some tensor calculation one will find that three non-conformal viscous terms should come from the transformation rule of $\sigma_{\langle \mu}^{~~\rho}\sigma_{\nu \rangle \rho}$, which are $P_{\mu\nu} \sigma_{\alpha \beta}^2$, $\sigma_{\mu\nu} \partial u$ and $P_{\mu\nu} (\partial u)^2$, not only $P_{\mu\nu} \sigma_{\alpha \beta}^2$. This is the reason that $\eta \tau_\pi^*$ and $\lambda_2$ of D4-brane that is got through KS proposal in \cite{Wu1807} do not match with the results got by direct calculation. We will see this in more detail in the following paragraphs. According to the analysis in this paragraph, we are of the opinion that the transformation rules for the viscous tensors in the dynamical sector should be
\begin{align}
  & \sideset{_\langle}{}{\mathop D} \sigma_{\mu\nu\rangle} \to \sideset{_\langle}{}{\mathop D} \sigma_{\mu\nu\rangle} + \chi P_{\mu\nu} D(\partial u) , \quad \sigma_{\mu\nu} \partial u \to \sigma_{\mu\nu}\partial u + \chi P_{\mu\nu}(\partial u)^2 , \quad \sigma_{\langle\mu}^{~~\rho} \Omega_{\nu\rangle\rho} \to \sigma_{\langle\mu}^{~~\rho} \Omega_{\nu\rangle\rho}, \cr
  & \sigma_{\langle\mu}^{~~\rho} \sigma_{\nu\rangle\rho} \to \sigma_{\langle\mu}^{~~\rho} \sigma_{\nu\rangle\rho} + \chi P_{\mu\nu} \sigma_{\alpha\beta}^2 + 2(d-1) \chi \sigma_{\mu\nu} \partial u + \chi \left(  \chi - \frac1{\tilde d-1} \right) P_{\mu\nu} (\partial u)^2.
\end{align}
Compared with \cite{Wu1807}, the rule for $\sigma_{\langle\mu}^{~~\rho} \sigma_{\nu\rangle\rho}$ has been changed.

Based on the new transformation rules, we get new relations between the lower dimensional non-conformal coefficients and the conformal ones in higher dimensions (the ones with tildes) as
\begin{align}\label{eq: relations of dynamical coef between conf and non-conf}
  \eta\tau_{\pi} &= \widetilde{\eta\tau_{\pi}}, \qquad \lambda_1 = \tilde{\lambda}_1, \qquad \lambda_2 = \tilde{\lambda}_2, \qquad \zeta\tau_{\Pi} = 2 \chi \widetilde{\eta\tau_{\pi}}, \qquad \xi_1 = \chi \tilde{\lambda}_1, \cr
  \eta\tau_{\pi}^* &= (d-1) \chi (4 \tilde \lambda_1 - \widetilde{\eta \tau_\pi}), \qquad \xi_2 = 4 \chi \left( \chi - \frac{1}{\tilde{d}-1} \right) \tilde \lambda_1 + \frac{2\chi}{\tilde{d}-1} \widetilde{\eta\tau_{\pi}}.
\end{align}
From the above, one can see that the conformal coefficients $\eta \tau_\pi,\; \lambda_1$ and $\lambda_2$ of the lower dimensional, non-conformal fluid inherit the values from their higher dimensional counterparts. While the non-conformal dynamical coefficients $\zeta \tau_\Pi$, $\xi_1$, $\eta \tau_\pi^*$, and $\xi_2$ are changed by the factor $\chi$ or by both $\chi$ and combinations of $\eta \tau_\pi$ and $\lambda_1$. We also know that the conformal coefficients do not contain $q$ whereas the non-conformal ones do. From \cref{eq: relations of dynamical coef between conf and non-conf}, we can conclude that the $q$-dependence of the non-conformal coefficients are actually through the factor $\chi$, which can be seen from \cref{eq: q-dependence of chi}.
The physical reason that the non-conformal dynamical coefficients are $q$-dependent has been pointed out in \cite{Wu2012} that the viscous terms of those coefficients all contain $\partial u$, which is the trace of $\partial_\mu u_\nu$. The dimensional reduction will diminish the components in taking the trace, causing their changes.

It has been found in \cite{Wu1807} that $\eta\tau_\pi^*$ and $\xi_2$ of D4-brane can not be got correctly. From \cref{eq: relations of dynamical coef between conf and non-conf} we can see that it is the identities involved with these two are not got correctly in \cite{Wu1807} since the contribution from $\tilde \lambda_1$ is missing. By the relations in \cref{eq: relations of dynamical coef between conf and non-conf}, one can check that this problem is fixed now. If one put the explicit expression of $\chi$ \cref{eq: q-dependence of chi} into the relations of $\eta \tau_\pi^*$ and $\xi_2$ in \cref{eq: relations of dynamical coef between conf and non-conf}. One will find that the identities of $\eta \tau_\pi^*$ and $\xi_2$ become exactly the Kleinert-Probst relations \cref{eq: Kleinert-Probst relations}. Thus the KS proposal is in nature the same as the Romatschke and Kleinert-Probst relations. The difference is that the KS proposal stresses the origins of the non-conformal viscous terms from the conformal viscous tensors, whereas the Romatschke and Kleinert-Probst relations offer directly the identities between non-conformal and the conformal coefficients.

\subsection{The thermodynamical sector}

In the KS proposal, the viscous tensors of the thermal sector will not change as in \cref{eq: substitution of thermal sector in KS proposal}. But we would like to make a different claim for the curvature tensor in the thermal sector that they will change in the dimensional reduction as
\begin{align}
  R_{\langle\mu\nu\rangle} &\to R_{\langle\mu\nu\rangle} + \chi P_{\mu\nu} R - (\tilde d-1) (d-1) \chi \, u^{\rho} u^{\sigma} R_{\rho\langle\mu\nu\rangle\sigma}, \\
  2 u^{\rho} u^\sigma R_{\rho\langle\mu\nu\rangle\sigma} &\to 2u^{\rho} u^\sigma R_{\rho\langle\mu\nu\rangle\sigma} - (\tilde d-1) \chi \, P_{\mu\nu} u^\rho u^\sigma R_{\rho\sigma}.
\end{align}

Then the term in the conformal stress-energy tensor of higher dimensions
\begin{align}
  \tilde \kappa \left( R_{\langle\mu\nu\rangle} - 2u^\rho u^\sigma R_{\rho\langle\mu\nu\rangle\sigma} \right)
\end{align}
will transform as
\begin{align}
  &\; \tilde \kappa \left( R_{\langle\mu\nu\rangle} - 2u^\rho u^\sigma R_{\rho\langle\mu\nu\rangle\sigma} \right) \cr
  \to&\; \tilde \kappa \left( R_{\langle\mu\nu\rangle} - 2 u^\rho u^\sigma R_{\rho\langle\mu\nu\rangle\sigma} \right) + \left( - \frac12 (\tilde d-1) (d-1) \chi \tilde \kappa \right) 2 u^\rho u^\sigma R_{\rho\langle\mu\nu\rangle\sigma} \cr
  & + P_{\mu\nu} \left( \chi \tilde \kappa R + (\tilde d-1) \chi \tilde \kappa u^\rho u^\sigma R_{\rho\sigma} \right) \cr
  =&\; \kappa \left( R_{\langle\mu\nu\rangle} - 2 u^\rho u^\sigma R_{\rho\langle\mu\nu\rangle\sigma} \right) + \kappa^* \cdot 2 u^\rho u^\sigma R_{\rho\langle\mu\nu\rangle\sigma} + P_{\mu\nu} \left( \xi_5 R + \xi_6 u^\rho u^\sigma R_{\rho\sigma} \right).
\end{align}
Thus one has
\begin{align}
  \kappa = \tilde \kappa, \qquad \kappa^* = - \frac12 (\tilde{d}-1) (d-1) \chi \tilde{\kappa}, \qquad \xi_5 = \chi \tilde{\kappa} , \qquad \xi_6 = (\tilde d-1) \chi \tilde{\kappa}. \label{eq: KS proposal for thermal sector}
\end{align}
The conformal coefficient $\kappa$ inherits the result of its higher-dimensional counterpart $\tilde \kappa$, the same as the other conformal coefficients. The non-conformal thermodynamical coefficients also get the $q$-dependence from $\chi$, the same as the non-conformal coefficients in the dynamical sector.

If one put \cref{eq: q-dependence of chi} into the identities of non-conformal coefficients in \cref{eq: KS proposal for thermal sector}, one will get the Romatschke relations for $\kappa^*$ and $\xi_{5,6}$ in general dimensions \cref{eq: relations of kappa* xi5 xi6 with kappa}. This again proves that the KS proposal has the same nature as the Romatschke and Kleinert-Probst relations.

At the last of this section, we would like to mention another observation about the algebraic identities of the second-order transport coefficients. The total number of them is 10, which can be listed as follows:
\begin{align}
  & 4\lambda_1 + \lambda_2 = 2 \eta \tau_\pi; \\
  & \tau_\Pi = \tau_\pi, \qquad \xi_1 = \frac{1}{p-q} [1 - (p-q) c_s^2] \lambda_1, \\
  & \eta\tau^*_\pi = \left(1 - (p-q) c_s^2 \right) (4\lambda_1 - \eta\tau_\pi), \cr
  & \xi_2 = \frac{2 \left(1 - (p-q) c_s^2 \right)}{(p-q)^2} \left[ (1 - 2(p-q) c_s^2) 2\lambda_1 + (p-q) c_s^2 \eta\tau_\pi \right]; \\
  & \kappa^* = - \frac{1-(p-q) c_s^2}{2 c_s^2} \kappa, \qquad \xi_5 = \frac{1-(p-q) c_s^2}{p-q} \kappa, \qquad \xi_6 = \frac{1-(p-q) c_s^2}{(p-q) c_s^2} \kappa; \\
  & \xi_3 = \frac{1 - (p-q) c_s^2}{p-q} \lambda_3, \qquad \xi_4 = - \frac{2}{p-q} \lambda_4.
\end{align}
Only the first one, i.e. the Haack-Yarom relation is purely among the conformal coefficients. The other 9 identities all speak the relations between the non-conformal coefficients and the conformal ones. Why is Haack-Yarom relation so special and what on earth does it speak of? We hope future studies can give us a satisfying answer.

\section{Discussions and outlook}

In this paper, we have proved that hitherto the gravity models that can be used to offer analytic results for the second-order transport coefficients of the non-conformal relativistic fluids are all Chamblin-Reall type. The only difference is the number of independent background scalar fields. Through previous studies and our works, we find that only the Chamblin-Reall models with one background scalar can be solved exactly. These models are the D(p-q)-brane and the compactified AdS black hole. We calculate all the 7 dynamical second-order transport coefficients for the compactified AdS black hole in this paper via the fluid/gravity correspondence.

We not only calculate the dynamical coefficients for the compactified AdS black hole, but also manage to get the thermodynamical coefficients through formal generalization. Thus we get non-trivial results for 11 second-order transport coefficients about compactified AdS black hole. The trivial ones are the $\lambda_{3,4}$ and $\xi_{3,4}$. The compactified AdS black holes is the first example that we know everything about up to the second order.

Since the (compactified) Dp-brane and the compactified AdS black hole are the only two examples of the Chamblin-Reall type backgrounds with one scalar. We believe that there are also 11 non-trivial second-order transport coefficients for (compactified) Dp-brane. We have already known the 7 dynamical ones. The coefficients that remain unknown in the thermal sector should also be $\kappa,\; \kappa^*$ and $\xi_{5,6}$, which may need the Minkowski-space AdS/CFT correspondence to derive.

We also find the physical meaning of the Kanitscheider-Skenderies proposal. This proposal recovers the origins of the non-conformal viscous terms from the conformal viscous tensors, through which one can get algebraic identities for the non-conformal coefficients in terms of the conformal ones. These identities are equivalent to the Romatschke and Kleinert-Probst relations. We also claim a possible way of change for the viscous tensors in the thermodynamical sector which have trivial reducing behaviors in the original KS proposal. It finally leads to the Romatschke relations for $\kappa^*$ and $\xi_{5,6}$ in general dimensions.

At last, we would like to talk a little about the study of the transport coefficients for NS5-brane, which has not been referred to in the works on (compactified) Dp-brane \cite{Wu1807,Wu2012}. It can be proved that NS5-brane and D5-brane are completely the same if written in Einstein frame. The first-order perturbations of D5-brane are found not renormalizable \cite{Wu1807}, it is not dual to any physical relativistic fluid in fluid/gravity correspondence, so is NS5-brane. Considering \cite{Parnachev0506} gets the bulk viscosity for NS5-brane with real-time AdS/CFT correspondence. Maybe we should use the same method to extract the second-order transport coefficients for NS5-brane.

\section*{Acknowledgement}

C. Wu would like to thank Si-wen Li for helpful discussions on the D(-1)-D3 background. The authors of this work would like to thank the Young Scientists Fund of the National Natural Science Foundation of China (Grant No. 11805002) for its support.

\appendix

\section{The reduction ansatz}

In the calculation of reducing the metric of AdS$_{p+2}$ black hole to a $(p-q+2)$-dimensional manifold, we use the ansatz
\begin{align}
	ds^2 = e^{2\alpha_1 A} g_{MN} dx^{M} dx^N + e^{2\alpha_2 A}\delta_{mn}dy^m dy^n.
\end{align}
From the above one can calculate the Christoffel symbol as
\begin{align}
  \widetilde\Gamma^M_{N P} &= \Gamma^M_{N P} + \alpha_1 ( \delta^M_N \partial_P A + \delta^M_P \partial_N A - g_{NP}\nabla^M A ), \cr
  \widetilde\Gamma^M_{mn} &= - \alpha_2 (\nabla^M A) e^{(- 2\alpha_1 + 2 \alpha_2) A} \delta_{mn}, \cr
  \widetilde\Gamma^n_{M m} &= \alpha_2 \partial_M A \delta^n_m.
\end{align}
Here the symbols with tilde are $(p+2)$-dimensional while those without are $(p-q+2)$-dimensional. One will also find the following summations of the Christoffel symbols are useful in the calculation, which are
\begin{align}
  \widetilde\Gamma^{N}_{M N} &= \Gamma^{N}_{M N} + (p-q+2) \alpha_1\partial_M A, \cr
  \widetilde\Gamma^{m}_{M m} &= q \alpha_2\partial_M A, \cr
  \widetilde\Gamma^{\hat N}_{M \hat N} &= \Gamma^{N}_{MN} + [(p-q+2) \alpha_1 + q \alpha_2 ]\partial_M A.
\end{align}
Then, the $(p+2)$-dimensional Ricci tensor $\mathcal R_{\hat M \hat N}$ can be got as
\begin{align}
  \mathcal R_{M N} =&\; R_{M N} - [(p-q)\alpha_1 + q\alpha_2] \nabla_M \nabla_N A - \alpha_1 g_{M N} \nabla^2 A \cr
  & + \left[ (p-q) \alpha_1^2 + 2q \alpha_1\alpha_2 - q \alpha_2^2 \right] \partial_M A \partial_N A  \cr
  & - \left[ (p-q) \alpha_1^2 + q \alpha_1\alpha_2 \right] g_{M N} (\partial A)^2,  \\
  \mathcal R_{mn} =& - \left[ \alpha_2 \nabla^2 A + \left( (p-q) \alpha_1 \alpha_2 + q \alpha_2^2 \right) (\partial A)^2 \right] e^{(- 2\alpha_1 + 2 \alpha_2) A} \delta_{mn},
\end{align}
with $R_{MN}$ the Ricci tensor in $p-q+2$ dimensions. Finally the Ricci scalar is calculated as
\begin{align}
  \mathcal R =&\; e^{-2\alpha_1 A} \left[ R - 2 ((p-q+1) \alpha_1 + q \alpha_2) \nabla^2 A - \left( (p-q)(p-q+1) \alpha_1^2 \right.\right.\cr
  &\left.\left. + 2q(p-q) \alpha_1\alpha_2 + q(q+1) \alpha_2^2 \right) (\partial A)^2\right].
\end{align}
Here $\mathcal R$ and $R$ are separately the Ricci scalars of AdS$_{p+2}$ black hole and the reduced $(p-q+2)$-dimensional spacetime.

\section{The D(-1)-D3 background}

The 10-dimensional action of D(-1)-D3 model in Einstein frame is
\begin{align}
  S =&\; \frac1{2\kappa_{10}^2} \int d^{10}x \sqrt{-G} \left[\mathcal R- \frac{1}{2}(\nabla \phi)^2 + \frac{g_s^2}{2} e^{2\phi} (\nabla \chi )^2 - \frac{g_s^2}{2\cdot 5!} F^2_{M_1\cdots M_5} \right] \cr
  &- \frac{1}{\kappa_{10}^2} \int d^{9}x \sqrt{-H} \mathcal K + \frac{1}{\kappa_{10}^2} \int d^{9}x \sqrt{-H} \frac3{L_3}.
\end{align}
The first, second and third terms in the above are respectively the bulk term, the Gibbons-Hawking term and the counter term. Here we do not employ the self-dual condition for $F^2_{M_1\cdots M_5}$ as the normalization constant is $\frac1{2\cdot 5!}$ but not $\frac1{4\cdot 5!}$. In order to describe the D-instanton, one needs to make the Wick rotation for $\chi$ as $\chi \rightarrow i\chi$ \cite{Gibbons9511}. That's the reason for the sign difference between the two background scalars. By the form of the counter term, we can see that this background is actually conformal, since it is the same as the AdS black hole's. A non-conformal background should have a counter term that contains the contribution of background scalars. This can be seen from the cases of (compactified) Dp-brane \cite{Mas0703,Wu1508,Wu1604,Wu1807,Wu2012} and compactified AdS black hole in this paper.

The background fields are the metric tensor, dilaton, axion and the Ramond-Ramond field:
\begin{align}\label{eq: D(-1)D(3) metric}
	ds^2 &= \left( \frac{r}{L_3} \right)^2 \left( - f(r) dt^2 + d \vv x^2 \right) + \left( \frac{L_3}{r} \right)^{2} \left( \frac{dr^2}{f(r)} + r^2 d\Omega_5^2 \right), \cr
    e^\phi &= H_{-1}, \qquad \chi = g_s^{-1} (1-H_{-1}^{-1}), \qquad F_{\theta_1\cdots\theta_5} = g_s^{-1} Q_3 \sqrt{\gamma_5}.
\end{align}
Here we have the blackening factor $f(r)=1 - \frac{r_H^4}{r^4}$ and the harmonic function for D-instanton $H_{-1} = 1 - \frac{L_{-1}^4}{r_H^4}\ln f$. $L_{-1}$ is a constant related with D-instanton and $Q_3 = 4L_3^4$. Using the reducing procedure as in \cite{Wu1807}, we get the 5-dimensional reduced action
\begin{align}
  S = \frac1{2\kappa_{5}^2} \int d^{5}x \sqrt{-g} \left[R- \frac{1}{2}(\partial \phi)^2 +\frac12 e^{2\phi}(\partial \chi )^2 + \frac{12}{L_3^2}\right] - \frac{1}{\kappa_{5}^2} \int d^{4}x \sqrt{-h} \left( K - \frac{3}{L_3} \right).
\end{align}
In the above, $\frac{1}{2\kappa_{5}^2} = \frac{L_3^5\Omega_5}{2\kappa_{10}^2}$ and we have set $g_s=1$. The reduced 5-dimensional metric is just AdS$_5$ black hole
\begin{align}\label{eq: D(-1)D(3) 5 dimensional reduced metric}
	ds^2 = \frac{r^2}{L_3^2} \left( -f(r) dt^2 + d \vv x^2 \right) + \frac{L_3^2}{r^2} \frac{dr^2}{f(r)}
\end{align}
coupled with two scalar fields:
\begin{align}
  e^\phi &= H_{-1}, \qquad \chi = 1 - H_{-1}^{-1} = 1 - e^{-\phi}.
\end{align}

\section{The transport coefficients of NS5-brane}

In this part of the appendix, we would like to point out the reason why fluid/gravity correspondence can not extract the transport coefficients for the NS5-brane. The answer is that in Einstein frame, NS5-brane is completely the same as D5-brane whose transport coefficients can not be extract in fluid/gravity correspondence. So we can not use the present method to study NS5-brane.

The 10-dimensional action and background solution in Einstein frame for NS5-brane can be written as
\begin{align}
  S &= \frac{1}{2\kappa_{10}^2} \int d^{10}x \sqrt{-g} \left[ R - \frac12 (\partial \phi)^2 - \frac{e^{-\phi}}{2\cdot 3!} \widetilde H_3^2 \right], \\
  ds^2 &= H^{-\frac14} (- f(r) dt^2 + d \vv x^2) + H^\frac34 \left( \frac{dr^2}{f(r)} + r^2 d\Omega_3^2 \right), \\
  e^\phi &= H^\frac12, \qquad \widetilde H_{\theta_1 \theta_2 \theta_3} = Q \sqrt{\gamma_3}, \qquad H = 1 + \frac{r_5^2}{r^2}, \qquad f = 1 - \frac{r_H^2}{r^2}.
\end{align}
Here the $\widetilde H_3$ is the NS-NS background field magnetically couples with the NS5-brane, $H$ is the harmonic function. For D5-brane one has
\begin{align}
  S &= \frac1{2\kappa_{10}^2} \int d^{10}x \sqrt{-g} \left[ R - \frac12 (\partial \phi)^2 - \frac{g_s^2}{2\cdot3!} e^\phi \widetilde F_3^2 \right], \\
  ds^2 &= - H_5^{-\frac14} \left( - f(r) dt^2 + d \vv x^2 \right) + H_5^\frac34 \left( \frac{dr^2}{f(r)} + r^2 d\Omega_3^2 \right), \\
  e^\phi &= H_5^{-\frac12}, \qquad \widetilde F_{\theta_1 \theta_2 \theta_3} = g_s^{-1} Q_5 \sqrt{\gamma_3}, \qquad H_5 = 1 + \frac{r_5^2}{r^2}, \qquad f = 1 - \frac{r_H^2}{r^2}
\end{align}
In the above, $\widetilde F_3$ is the Ramond-Ramond field that couples with the D5-brane, and the $H_5$ is the harmonic function for D5-brane.

The only difference in the actions of NS5- and D5-brane is how dilaton couples with the 3-form fields. For NS5-brane it is $e^{-\phi}$ whereas for D5-brane is $e^\phi$. But as we can see from the profiles of the dilatons for these two kinds of branes that NS5-brane is $e^\phi = H^\frac12$ and D5-brane is $e^\phi = H_5^{-\frac12}$. Thus the terms that coupled with the 3-form fields for these two cases are actually the same. So both the action and the metric for NS5- and D5-brane are the same.

As what has been calculated and discussed in \cite{Wu1807}, the D5-brane is not dual to any physical relativistic fluid in the framework of fluid/gravity correspondence. So we can not extract any useful information for NS5-brane either, with the present means.


\begin{thebibliography}{10}

\bibitem{Wu2012}
C.~Wu and Y.~Wang, {\it {Second order transport coefficients of nonconformal
  fluids from compactified Dp-branes}},  {\em JHEP} {\bf 05} (2021) 262,
  [\href{http://arxiv.org/abs/2012.14699}{{\tt arXiv:2012.14699}}].

\bibitem{Son0205}
D.~T. Son and A.~O. Starinets, {\it {Minkowski space correlators in AdS/CFT
  correspondence: Recipe and applications}},  {\em JHEP} {\bf 09} (2002) 042,
  [\href{http://arxiv.org/abs/hep-th/0205051}{{\tt hep-th/0205051}}].

\bibitem{Policastro0205}
G.~Policastro, D.~T. Son, and A.~O. Starinets, {\it {From AdS / CFT
  correspondence to hydrodynamics}},  {\em JHEP} {\bf 09} (2002) 043,
  [\href{http://arxiv.org/abs/hep-th/0205052}{{\tt hep-th/0205052}}].

\bibitem{Policastro0210}
G.~Policastro, D.~T. Son, and A.~O. Starinets, {\it {From AdS / CFT
  correspondence to hydrodynamics. 2. Sound waves}},  {\em JHEP} {\bf 12}
  (2002) 054, [\href{http://arxiv.org/abs/hep-th/0210220}{{\tt
  hep-th/0210220}}].

\bibitem{Baier0712}
R.~Baier, P.~Romatschke, D.~T. Son, A.~O. Starinets, and M.~A. Stephanov, {\it
  {Relativistic viscous hydrodynamics, conformal invariance, and holography}},
  {\em JHEP} {\bf 04} (2008) 100, [\href{http://arxiv.org/abs/0712.2451}{{\tt
  arXiv:0712.2451}}].

\bibitem{Arnold1105}
P.~Arnold, D.~Vaman, C.~Wu, and W.~Xiao, {\it {Second order hydrodynamic
  coefficients from 3-point stress tensor correlators via AdS/CFT}},  {\em
  JHEP} {\bf 10} (2011) 033, [\href{http://arxiv.org/abs/1105.4645}{{\tt
  arXiv:1105.4645}}].

\bibitem{Bhattacharyya0712}
S.~Bhattacharyya, V.~E. Hubeny, S.~Minwalla, and M.~Rangamani, {\it {Nonlinear
  Fluid Dynamics from Gravity}},  {\em JHEP} {\bf 02} (2008) 045,
  [\href{http://arxiv.org/abs/0712.2456}{{\tt arXiv:0712.2456}}].

\bibitem{Bhattacharyya0803}
S.~Bhattacharyya, V.~E. Hubeny, R.~Loganayagam, G.~Mandal, S.~Minwalla,
  T.~Morita, M.~Rangamani, and H.~S. Reall, {\it {Local Fluid Dynamical Entropy
  from Gravity}},  {\em JHEP} {\bf 06} (2008) 055,
  [\href{http://arxiv.org/abs/0803.2526}{{\tt arXiv:0803.2526}}].

\bibitem{Gubser0804PRL}
S.~S. Gubser, A.~Nellore, S.~S. Pufu, and F.~D. Rocha, {\it {Thermodynamics and
  bulk viscosity of approximate black hole duals to finite temperature quantum
  chromodynamics}},  {\em Phys. Rev. Lett.} {\bf 101} (2008) 131601,
  [\href{http://arxiv.org/abs/0804.1950}{{\tt arXiv:0804.1950}}].

\bibitem{Gubser0804PRD}
S.~S. Gubser and A.~Nellore, {\it {Mimicking the QCD equation of state with a
  dual black hole}},  {\em Phys. Rev. D} {\bf 78} (2008) 086007,
  [\href{http://arxiv.org/abs/0804.0434}{{\tt arXiv:0804.0434}}].

\bibitem{Gubser0806}
S.~S. Gubser, S.~S. Pufu, and F.~D. Rocha, {\it {Bulk viscosity of strongly
  coupled plasmas with holographic duals}},  {\em JHEP} {\bf 08} (2008) 085,
  [\href{http://arxiv.org/abs/0806.0407}{{\tt arXiv:0806.0407}}].

\bibitem{Finazzo1412}
S.~I. Finazzo, R.~Rougemont, H.~Marrochio, and J.~Noronha, {\it {Hydrodynamic
  transport coefficients for the non-conformal quark-gluon plasma from
  holography}},  {\em JHEP} {\bf 02} (2015) 051,
  [\href{http://arxiv.org/abs/1412.2968}{{\tt arXiv:1412.2968}}].

\bibitem{Attems1603}
M.~Attems, J.~Casalderrey-Solana, D.~Mateos, I.~Papadimitriou,
  D.~Santos-Oliv\'an, C.~F. Sopuerta, M.~Triana, and M.~Zilh\~ao, {\it
  {Thermodynamics, transport and relaxation in non-conformal theories}},  {\em
  JHEP} {\bf 10} (2016) 155, [\href{http://arxiv.org/abs/1603.01254}{{\tt
  arXiv:1603.01254}}].

\bibitem{Kleinert1610}
P.~Kleinert and J.~Probst, {\it {Second-Order Hydrodynamics and Universality in
  Non-Conformal Holographic Fluids}},  {\em JHEP} {\bf 12} (2016) 091,
  [\href{http://arxiv.org/abs/1610.01081}{{\tt arXiv:1610.01081}}].

\bibitem{Li1411}
D.~Li, S.~He, and M.~Huang, {\it {Temperature dependent transport coefficients
  in a dynamical holographic QCD model}},  {\em JHEP} {\bf 06} (2015) 046,
  [\href{http://arxiv.org/abs/1411.5332}{{\tt arXiv:1411.5332}}].

\bibitem{Buchel0311}
A.~Buchel and J.~T. Liu, {\it {Universality of the shear viscosity in
  supergravity}},  {\em Phys. Rev. Lett.} {\bf 93} (2004) 090602,
  [\href{http://arxiv.org/abs/hep-th/0311175}{{\tt hep-th/0311175}}].

\bibitem{Buchel0406200}
A.~Buchel, {\it {N=2* hydrodynamics}},  {\em Nucl. Phys.} {\bf B708} (2005)
  451--466, [\href{http://arxiv.org/abs/hep-th/0406200}{{\tt hep-th/0406200}}].

\bibitem{Benincasa0507}
P.~Benincasa, A.~Buchel, and A.~O. Starinets, {\it {Sound waves in strongly
  coupled non-conformal gauge theory plasma}},  {\em Nucl. Phys.} {\bf B733}
  (2006) 160--187, [\href{http://arxiv.org/abs/hep-th/0507026}{{\tt
  hep-th/0507026}}].

\bibitem{Buchel0812}
A.~Buchel and C.~Pagnutti, {\it {Bulk viscosity of N=2* plasma}},  {\em Nucl.
  Phys.} {\bf B816} (2009) 62--72, [\href{http://arxiv.org/abs/0812.3623}{{\tt
  arXiv:0812.3623}}].

\bibitem{Buchel0908}
A.~Buchel, {\it {Relaxation time of non-conformal plasma}},  {\em Phys. Lett.
  B} {\bf 681} (2009) 200--203, [\href{http://arxiv.org/abs/0908.0108}{{\tt
  arXiv:0908.0108}}].

\bibitem{Buchel1110}
A.~Buchel, {\it {Violation of the holographic bulk viscosity bound}},  {\em
  Phys. Rev. D} {\bf 85} (2012) 066004,
  [\href{http://arxiv.org/abs/1110.0063}{{\tt arXiv:1110.0063}}].

\bibitem{Buchel1503}
A.~Buchel, M.~P. Heller, and R.~C. Myers, {\it {Equilibration rates in a
  strongly coupled nonconformal quark-gluon plasma}},  {\em Phys. Rev. Lett.}
  {\bf 114} (2015), no.~25 251601, [\href{http://arxiv.org/abs/1503.07114}{{\tt
  arXiv:1503.07114}}].

\bibitem{Bigazzi0909}
F.~Bigazzi, A.~L. Cotrone, J.~Mas, A.~Paredes, A.~V. Ramallo, and J.~Tarrio,
  {\it {D3-D7 Quark-Gluon Plasmas}},  {\em JHEP} {\bf 11} (2009) 117,
  [\href{http://arxiv.org/abs/0909.2865}{{\tt arXiv:0909.2865}}].

\bibitem{Bigazzi0912}
F.~Bigazzi, A.~L. Cotrone, and J.~Tarrio, {\it {Hydrodynamics of fundamental
  matter}},  {\em JHEP} {\bf 02} (2010) 083,
  [\href{http://arxiv.org/abs/0912.3256}{{\tt arXiv:0912.3256}}].

\bibitem{Buchel0509}
A.~Buchel, {\it {Transport properties of cascading gauge theories}},  {\em
  Phys. Rev.} {\bf D72} (2005) 106002,
  [\href{http://arxiv.org/abs/hep-th/0509083}{{\tt hep-th/0509083}}].

\bibitem{Buchel0903}
A.~Buchel, {\it {Hydrodynamics of the cascading plasma}},  {\em Nucl. Phys. B}
  {\bf 820} (2009) 385--416, [\href{http://arxiv.org/abs/0903.3605}{{\tt
  arXiv:0903.3605}}].

\bibitem{Buchel0708}
A.~Buchel, {\it {Bulk viscosity of gauge theory plasma at strong coupling}},
  {\em Phys. Lett.} {\bf B663} (2008) 286--289,
  [\href{http://arxiv.org/abs/0708.3459}{{\tt arXiv:0708.3459}}].

\bibitem{Bigazzi1006}
F.~Bigazzi and A.~L. Cotrone, {\it {An elementary stringy estimate of transport
  coefficients of large temperature QCD}},  {\em JHEP} {\bf 08} (2010) 128,
  [\href{http://arxiv.org/abs/1006.4634}{{\tt arXiv:1006.4634}}].

\bibitem{Chamblin9903}
H.~A. Chamblin and H.~S. Reall, {\it {Dynamic dilatonic domain walls}},  {\em
  Nucl. Phys. B} {\bf 562} (1999) 133--157,
  [\href{http://arxiv.org/abs/hep-th/9903225}{{\tt hep-th/9903225}}].

\bibitem{Parnachev0506}
A.~Parnachev and A.~Starinets, {\it {The Silence of the little strings}},  {\em
  JHEP} {\bf 10} (2005) 027, [\href{http://arxiv.org/abs/hep-th/0506144}{{\tt
  hep-th/0506144}}].

\bibitem{Benincasa0605}
P.~Benincasa and A.~Buchel, {\it {Hydrodynamics of Sakai-Sugimoto model in the
  quenched approximation}},  {\em Phys. Lett.} {\bf B640} (2006) 108--115,
  [\href{http://arxiv.org/abs/hep-th/0605076}{{\tt hep-th/0605076}}].

\bibitem{Mas0703}
J.~Mas and J.~Tarrio, {\it {Hydrodynamics from the Dp-brane}},  {\em JHEP} {\bf
  05} (2007) 036, [\href{http://arxiv.org/abs/hep-th/0703093}{{\tt
  hep-th/0703093}}].

\bibitem{Natsuume0807}
M.~Natsuume, {\it {Causal hydrodynamics and the membrane paradigm}},  {\em
  Phys. Rev.} {\bf D78} (2008) 066010,
  [\href{http://arxiv.org/abs/0807.1392}{{\tt arXiv:0807.1392}}].

\bibitem{Springer0810}
T.~Springer, {\it {Sound Mode Hydrodynamics from Bulk Scalar Fields}},  {\em
  Phys. Rev.} {\bf D79} (2009) 046003,
  [\href{http://arxiv.org/abs/0810.4354}{{\tt arXiv:0810.4354}}].

\bibitem{Springer0902}
T.~Springer, {\it {Second order hydrodynamics for a special class of gravity
  duals}},  {\em Phys. Rev.} {\bf D79} (2009) 086003,
  [\href{http://arxiv.org/abs/0902.2566}{{\tt arXiv:0902.2566}}].

\bibitem{Kanitscheider0901}
I.~Kanitscheider and K.~Skenderis, {\it {Universal hydrodynamics of
  non-conformal branes}},  {\em JHEP} {\bf 04} (2009) 062,
  [\href{http://arxiv.org/abs/0901.1487}{{\tt arXiv:0901.1487}}].

\bibitem{David0901}
J.~R. David, M.~Mahato, and S.~R. Wadia, {\it {Hydrodynamics from the
  D1-brane}},  {\em JHEP} {\bf 04} (2009) 042,
  [\href{http://arxiv.org/abs/0901.2013}{{\tt arXiv:0901.2013}}].

\bibitem{Wu1508}
C.~Wu, Y.~Chen, and M.~Huang, {\it {Fluid/gravity correspondence: A
  nonconformal realization in compactified D4 branes}},  {\em Phys. Rev.} {\bf
  D93} (2016), no.~6 066005, [\href{http://arxiv.org/abs/1508.04038}{{\tt
  arXiv:1508.04038}}].

\bibitem{Wu1604}
C.~Wu, Y.~Chen, and M.~Huang, {\it {Fluid/gravity correspondence: Second order
  transport coefficients in compactified D4-branes}},  {\em JHEP} {\bf 01}
  (2017) 118, [\href{http://arxiv.org/abs/1604.07765}{{\tt arXiv:1604.07765}}].

\bibitem{Wu1608}
C.~Wu, Y.~Chen, and M.~Huang, {\it {Chiral vortical effect from the
  compactified D4-branes with smeared D0-brane charge}},  {\em JHEP} {\bf 03}
  (2017) 082, [\href{http://arxiv.org/abs/1608.04922}{{\tt arXiv:1608.04922}}].

\bibitem{Wu1807}
C.~Wu, {\it {Second order transport coefficients of nonconformal relativistic
  fluids in various dimensions from Dp-brane}},  {\em JHEP} {\bf 01} (2019)
  097, [\href{http://arxiv.org/abs/1807.08268}{{\tt arXiv:1807.08268}}].

\bibitem{Gibbons9511}
G.~W. Gibbons, M.~B. Green, and M.~J. Perry, {\it {Instantons and seven-branes
  in type IIB superstring theory}},  {\em Phys. Lett. B} {\bf 370} (1996)
  37--44, [\href{http://arxiv.org/abs/hep-th/9511080}{{\tt hep-th/9511080}}].

\bibitem{Greene1990NPB}
B.~R. Greene, A.~D. Shapere, C.~Vafa, and S.-T. Yau, {\it {Stringy Cosmic
  Strings and Noncompact Calabi-Yau Manifolds}},  {\em Nucl. Phys. B} {\bf 337}
  (1990) 1--36.

\bibitem{Aharony0506}
O.~Aharony, A.~Buchel, and A.~Yarom, {\it {Holographic renormalization of
  cascading gauge theories}},  {\em Phys. Rev. D} {\bf 72} (2005) 066003,
  [\href{http://arxiv.org/abs/hep-th/0506002}{{\tt hep-th/0506002}}].

\bibitem{Haack0806}
M.~Haack and A.~Yarom, {\it {Nonlinear viscous hydrodynamics in various
  dimensions using AdS/CFT}},  {\em JHEP} {\bf 10} (2008) 063,
  [\href{http://arxiv.org/abs/0806.4602}{{\tt arXiv:0806.4602}}].

\bibitem{Barnes1004}
E.~Barnes, D.~Vaman, C.~Wu, and P.~Arnold, {\it {Real-time finite-temperature
  correlators from AdS/CFT}},  {\em Phys. Rev.} {\bf D82} (2010) 025019,
  [\href{http://arxiv.org/abs/1004.1179}{{\tt arXiv:1004.1179}}].

\bibitem{Romatschke0906}
P.~Romatschke, {\it {Relativistic Viscous Fluid Dynamics and Non-Equilibrium
  Entropy}},  {\em Class. Quant. Grav.} {\bf 27} (2010) 025006,
  [\href{http://arxiv.org/abs/0906.4787}{{\tt arXiv:0906.4787}}].

\bibitem{Bhattacharyya1201}
S.~Bhattacharyya, {\it {Constraints on the second order transport coefficients
  of an uncharged fluid}},  {\em JHEP} {\bf 07} (2012) 104,
  [\href{http://arxiv.org/abs/1201.4654}{{\tt arXiv:1201.4654}}].

\bibitem{Moore1210}
G.~D. Moore and K.~A. Sohrabi, {\it {Thermodynamical second-order hydrodynamic
  coefficients}},  {\em JHEP} {\bf 11} (2012) 148,
  [\href{http://arxiv.org/abs/1210.3340}{{\tt arXiv:1210.3340}}].

\end{thebibliography}

\providecommand{\href}[2]{#2}\begingroup\raggedright\endgroup

\end{document}